\documentclass[11pt,prd,nofootinbib,reprint,superscriptaddress,longbibliography]{revtex4-2}

\usepackage{amsmath, amssymb, amsthm, graphicx, epsfig, fancyhdr,epsfig,multirow}
\usepackage[utf8]{inputenc}
\usepackage{bbold}
\usepackage{slashed}
\usepackage{tikz-feynman}
\usepackage{amsmath}
\usepackage{amsfonts}
\usepackage{amssymb}
\usepackage{xcolor}
\usepackage{comment}
\usepackage{subcaption}
\usepackage[normalem]{ulem}

\tikzfeynmanset{compat=1.1.0}

\usepackage{tikzsymbols}
\usepackage{natbib}
\usepackage{hyperref}

\makeatletter
\def\alt{\mathrel{\mathpalette\gl@align<}}
\def\agt{\mathrel{\mathpalette\gl@align>}}
\def\gl@align#1#2{\lower.6ex\vbox{\baselineskip\z@skip\lineskip\z@
\ialign{$\m@th#1\hfil##\hfil$\crcr#2\crcr\sim\crcr}}} \makeatother

\def\beq{\begin{equation}}
\def\eeq{\end{equation}}
\def\bea{\begin{eqnarray}}
\def\eea{\end{eqnarray}}
\def\nn{\nonumber}

\DeclareUnicodeCharacter{2212}{-}

\newcommand{\rk}{R_K}
\newcommand{\rks}{R_{K^\ast}}
\newcommand{\rkrks}{R_{K^{(\ast)}}}
\newcommand{\rdrds}{R_{D^{(\ast)}}}

\pdfoutput=1

\begin{document}

\pagestyle{plain}

\title{$\rkrks$ from RPV-SUSY sneutrinos}
\author{Debjyoti Bardhan}
\email{bardhan@post.bgu.ac.il}
\affiliation{Department of Physics, Ben-Gurion University of the Negev, Israel}
\affiliation{Department of Physics, Indian Institute of Science Education and Research Pune, India}

\author{Diptimoy Ghosh}
\email{diptimoy.ghosh@iiserpune.ac.in}
\affiliation{Department of Physics, Indian Institute of Science Education and Research Pune, India}

\author{Divya Sachdeva}
\email{divya.sachdeva@students.iiserpune.ac.in}

\affiliation{Department of Physics, Indian Institute of Science Education and Research Pune, India}
\affiliation{Laboratoire de Physique Theorique et Hautes Energies (LPTHE), Paris, France}

\begin{abstract}
We analyze the lepton flavor universality violation (LFV) in process $b\to s \ell^+\ell^-$ in an R-parity violating supersymmetric (RPV-SUSY) scenario. The most recent update on $\rk$ from the LHCb collaboration suggests approximately $3.1\sigma$ deviation from the Standard Model predictions, strengthening the case of LFV New Physics. In this work, we show that $R_{K^{(*)}}$ anomaly can be addressed via only sneutrinos within the framework of R-parity violating interactions assuming phenomenologically viable values of the couplings.  While our proposed solution is by no means the only solution to these anomalies, it is a phenomenologically plausible one and, notably, quite a minimal one in the context of RPV-SUSY. 
\end{abstract}
\maketitle
\section{Introduction}
Gauge invariance of the Standard Model (SM) predicts lepton flavour universality (LFU), i.e. the couplings of leptons to electroweak bosons are independent of the generation the leptons belong to. Given two similar electroweak scattering or decay processes involving leptons from different generations, the entire difference in the branching ratios between the two processes can be attributed to the difference in the masses of the leptons. While this has been stringently verified by leptonic decays of the $W$ and $Z$ bosons, both neutral and charged current decays of B-mesons hint at violation of LFU. This can be probed by measuring the branching ratios, or rather, the ratio of the branching ratios, of these decays. For charged current decays of the $B$-meson, the ratios $R_{D^{(\ast)}} = \frac{\mathcal{B}(B \to D^{(\ast)} \tau \nu)}{\mathcal{B}(B \to D^{(\ast)} \ell \nu)}, (\ell = e,\mu)$ have been widely studied \cite{Fajfer:2012vx,Freytsis:2015qca,
Choudhury:2016ulr,Choudhury:2012hn,Bardhan:2016uhr, Azatov:2018knx,Hu:2018veh,Asadi:2019xrc,Shi:2019gxi,Becirevic:2019tpx,Gomez:2019xfw,
Asadi:2020fdo,Bernlochner:2017jka,
Jung:2018lfu,Choudhury:2017qyt,Greljo:2018ogz,
Feruglio:2018fxo,Hu:2018lmk,Huang:2018nnq,Bardhan:2019ljo}. The similar ratios for the neutral current decays of the $B$-meson that have been studied are $\rkrks = \frac{\mathcal{B}(B \to K^{(\ast)} \mu^+ \mu^-)}{\mathcal{B}(B \to K^{(\ast)} e^+ e^-)}$ 
\cite{Ghosh:2014awa,Bhattacharya:2016mcc,
Greljo:2015mma,Bardhan:2017xcc, Ghosh:2017ber,Biswas:2020uaq,Datta:2019bzu,
DelleRose:2019ukt,Alok:2019ufo,
Ciuchini:2019usw,Datta:2019tuj,
Bhattacharya:2019eji, Trifinopoulos:2018rna, Trifinopoulos:2019lyo,
Wang:2017mrd,DAmbrosio:2017wis,
Cornella:2018tfd,Altmannshofer:2017yso,
Falkowski:2015zwa,Alonso:2015sja,Mandal:2014kma,Alda:2020okk,
Saad:2020ucl,Li:2018rax,Azatov:2018kzb,Ciuchini:2020gvn,DAmico:2017mtc,Azatov:2018kzb,Cornella:2021sby}. Measurements of both these ratios have shown them to be different from the value expected from the SM and this has led to a lot of work trying to resolve
the discrepancy using New Physics (NP) models (See Refs. \cite{Davighi:2021oel,Altmannshofer:2021qrr, Du:2021zkq, Marzocca:2021azj, Ban:2021tos, Lancierini:2021sdf, Hiller:2021pul,Alguero:2021anc,Lancierini:2021sdf} for a short list of papers that appeared after the Moriond 2021 Conference).

The value predicted by the SM is $\rk = 1.00 \pm 0.01$ \cite{Bordone:2016gaq}. A recent update from the {\sc LHCb } collaboration, $\rk^{\rm LHCb(2021)} = 0.846 ^{+0.044}_{-0.041}$ for the dilepton invariant mass-squared $q^2 \in [1.1, 6.0] \ {\rm GeV}^2$, has strengthened the existing discrepancy, which now stands at $3.1 \sigma$~\cite{Angelescu:2021lln}. This is updated from the earlier measurement which had the same central value but with a larger statistical error, in line with a smaller amount of data available for the previous measurement. The 2021 measurement has smaller uncertainty compared to the 2019 measurement which  suggests that the signal could be more than a statistical fluctuation.   Measurements of $\rk$ are summarized in Table~\ref{tab:rk_result} 

\begin{table}[h]
\begin{tabular}{|c|c|c|c|}
\hline
Collaboration & $q^2$ Bin (${\rm GeV}^2)$& Measurement & Ref.\\ 
\hline
{\sc BaBar (2012)} & $0.1 - 8.12$ & $0.74^{+0.40}_{-0.31} \pm 0.06$ & {\cite{Lees:2012tva}} \\
\hline
 {\sc Belle} (2020) & $1.0 - 6.0$ & $1.03^{+0.28}_{-0.24}\pm 0.01$ & \cite{Choudhury_2021} \\
 \hline
 {\sc LHCb} (2014) &  & $0.745^{+0.090}_{-0.074} \pm 0.036$ & \cite{Aaij:2014ora} \\ 
 {\sc LHCb} (2019) & $1.1 - 6.0$ & $0.846^{+0.060}_{-0.054} {}^{+0.016}_{-0.014}$ & \cite{Aaij:2019wad}\\
 {\sc LHCb} (2021) &  & $0.846^{+ 0.042}_{-0.039}{}^{+ 0.013}_{-0.012}$ & \cite{Aaij:2021vac} \\
 \hline
\end{tabular}
\caption{All measurements of $\rk$ made by different collaborations so far. The results are binned according to the value of $q^2$, which is the invariant mass of the dilepton system. The data used by the LHCb collaboration is the integrated luminosity available to it till that point.}
\label{tab:rk_result}
\end{table} 

Several phenomenological models have been used to explain the discrepancies, e.g. heavy leptoquarks~\cite{Angelescu:2021lln, Bauer:2015knc, Bigaran:2019bqv, Babu:2020hun, Alonso:2015sja,Barbieri:2015yvd,Sahoo:2016pet,Barbieri:2016las,Biggio:2016wyy,Datta:2019tuj,Crivellin:2021egp,Crivellin:2020mjs,Calibbi:2017qbu, Barbieri:2016las,Gripaios:2014tna,Das:2016vkr,
deMedeirosVarzielas:2018bcy,Sahoo:2018ffv,Bauer:2015knc, Becirevic:2017jtw,
Hiller:2017bzc,DiLuzio:2017vat,Calibbi:2015kma,Crivellin:2017zlb,Calibbi_2018,Blanke:2018sro, Carvunis:2021dss, Saad:2020ihm} and $Z^\prime$ models, including SUSY-$Z^\prime$ models, (See Refs~\cite{Gauld:2013qja,Buras:2013dea,Altmannshofer:2014cfa,Crivellin:2015mga,Greljo:2015mma,Falkowski:2015zwa,Chiang:2016qov,Allanach:2015gkd,Boucenna:2016qad,Datta:2017pfz,Crivellin:2015era, Alonso:2017uky,Alonso:2017bff,Duan:2018akc} for a partial list). Our approach in this note will be to explain the anomalies using a simplified model of R-parity violating supersymmetry (RPV-SUSY) with the minimal field content i.e via sneutrinos only. Our explanation of the $\rkrks$ anomaly assumes the suppression of the branching ratio of the $B \to K^{(*)} \mu^+ \mu^-$ channel, while leaving the electron channel untouched. This is motivated by the fact that some discrepancies in angular observables like $P_5^\prime$ can be explained by the alteration of the muonic channel. Excellent fit to all relevant observables is obtained by assuming $C_9^{\mu} = - C_{10}^{\mu}$ \cite{Alguero:2021anc, Hurth:2021nsi, Geng:2021nhg}. Earlier, RPV-SUSY was used to explain $B$-anomalies with different field content~\cite{Biswas:2014gga, Nandi:2006qe,Deshpand:2016cpw,Das:2017kfo,Altmannshofer:2017poe,Earl:2018snx,Hu:2019ahp,Altmannshofer:2020axr, Dev:2021zty}. For example, the authors in Ref.~\cite{Deshpand:2016cpw} attempt to explain the anomaly via one-loop contributions involving right-handed down type squarks ($\tilde{d_R}$). In Ref.~\cite{Das:2017kfo}, the authors consider the contribution to $b\to s\mu^+\mu^-$ transition from the box diagrams with $\tilde{d_R}$ and, a left-handed up type squark and sneutrino in the loop. In Ref.~\cite{Earl:2018snx}, the authors focus on parameters for which diagrams involving winos give large contributions whereas Ref.~\cite{Hu:2019ahp} consider diagrams involving winos only with sneutrinos. The authors in ref.~\cite{Altmannshofer:2020axr} explain anomaly via the diagrams involving third generation superpartners. Furthermore, it has to be noted that the RPV-SUSY scenario has enough room to accommodate a simultaneous explanation of the $R_{K^{(\ast)}}$, $R_{D^{(\ast)}}$ and the muon g-2 anomalies. This has been shown in Refs.~\cite{Altmannshofer:2020axr,Chen:2017hir,Zheng:2021wnu}. In the present work, we try to address $R_{K^{(\ast)}}$ anomalies via the sneutrino alone, which has not been explored earlier. We demonstrate that this scenario is phenomenologically viable if certain assumptions are made. We would like to 
stress here that, in the present work, we do not attempt to explain $\rkrks$, 
$\rdrds$ and $g-2$ anomalies simultaneously; instead, we just concentrate
on the explanation of $\rkrks$. 

The paper is arranged as follows: In Section~\ref{sec:Model_setup}, we provide a short description of the RPV-SUSY model we intend to use. Then, in Section~\ref{sec:bsmumu}, we attempt an explanation of the
discrepancy, mentioning all the assumptions we make to reach our goal. Finally, we conclude. 

\section{Model Setup}
\label{sec:Model_setup}
The well known RPV superpotential in terms of superfields is given by : 
\begin{equation}
W_{\mathrm{RPV}} \supset \frac{1}{2} \hat{\lambda}_{i j k} \hat{L}_{i} \hat{L}_{j} \hat{E}^c_{k}+\hat{\lambda}_{i j k}^{\prime} \hat{L}_{i} \hat{Q}_{j} \hat{D}^c_{k}+\frac{1}{2} \hat{\lambda}_{i j k}^{\prime \prime} \hat{U}_{i} \hat{D}_{j} \hat{D}_{k}
\end{equation}
where the hatted Latin letters denote superfields and the hatted Greek letters denote couplings. Note that $\hat{L}, \hat{Q}$ are left-handed lepton and quark chiral superfield doublets respectively, while $\hat{E}, \hat{U}, \hat{D}$ are right-handed electron, up and down quark chiral superfield singlets. Also note that, due to the anti-symmetric $SU(2)$ product in SUSY, the $\lambda$ coupling is anti-symmetric in its first two indices. 

 Since proton decay posits a very stringent constraint on the RPV couplings, we have to set either the lepton-number violating couplings ($\hat{\lambda}'$) or the baryon-number violating couplings ($\hat{\lambda}''$) to zero~\cite{Bhattacharyya:1998dt}. In order to explain the $\rkrks$ anomaly, we keep $\hat{\lambda}'$ non-zero and set $\hat{\lambda}''$ to zero. Thus, the superpotential simplifies to contain only the $\lambda$ and $\lambda^\prime$ terms. We assume that all the coloured SUSY particles -- the squarks and gluino -- are very heavy, as are the wino and Higgsino. Only the slepton doublet is assumed to be light enough to contribute to our process and the sneutrinos are taken to be degenerate in masses. Thus, we can write the RPV Lagrangian in terms of ordinary fields comprising SM and SUSY particles.

\begin{eqnarray}
\label{eqn:rpv_Lag}
\mathcal{L}_{\rm RPV} &\supset& -\frac{1}{2} \lambda_{i j k}\left[\left(\tilde{\nu}_{i L} \bar{l}_{k R} l_{j L}+\tilde{l}_{j L} \bar{l}_{k R} \nu_{i L}+\tilde{l}_{k R}^{*} \bar{\nu}_{i R}^{c} l_{j L}\right)\right. \nn \\
&& \left.  -\left(\tilde{\nu}_{j L} \bar{l}_{k R} l_{i L}+\tilde{l}_{i L} \bar{l}_{k R} \nu_{j L}+\tilde{l}_{k R}^{*} \bar{\nu}_{j R}^{c} l_{i L}\right) \right]  \\ 
&& -\lambda_{p q r}^{\prime}\left[\tilde{\nu}_{p L} \bar{d}_{r R} d_{q L}-V_{q l}^{\mathrm{CKM}}\tilde{l}_{p L} \bar{d}_{r R} u_{l L} \right] + {\rm H.c.}  \nn
\end{eqnarray}
The fields with a tilde above them are SUSY fields, while the rest are SM fields. Note that couplings are defined in mass basis in Eqn.~\ref{eqn:rpv_Lag}. For reviews on the RPV-SUSY, see Refs.~\cite{Barger:1989rk,Godbole:1992fb,Bhattacharyya:1996nj,Godbole:1999ye,Bhattacharyya:1997vv,Barbier:2004ez}.

Here, we  present a scenario where we keep $\lambda$ and $\lambda'$ both term non-zero and we found that keeping only sneutrino to be light can also address the $\rkrks$ anomaly. We could have chosen to keep other particles light and do a scan to find the correct parameter space. But, this has been considered in Refs.~\cite{Altmannshofer:2020axr}. So, in this work, we focus on the part of parameter space where only sneutrinos explain the anomaly. This is different from Ref~\cite{Hu:2019ahp} where only $\lambda'$ couplings were considered and thus, winos and sneutrinos were both required to be light.

\section{$b\to s\mu^+\mu^-$ process}
\label{sec:bsmumu}
\begin{figure}[!h]
\centering
\includegraphics[scale=0.15]{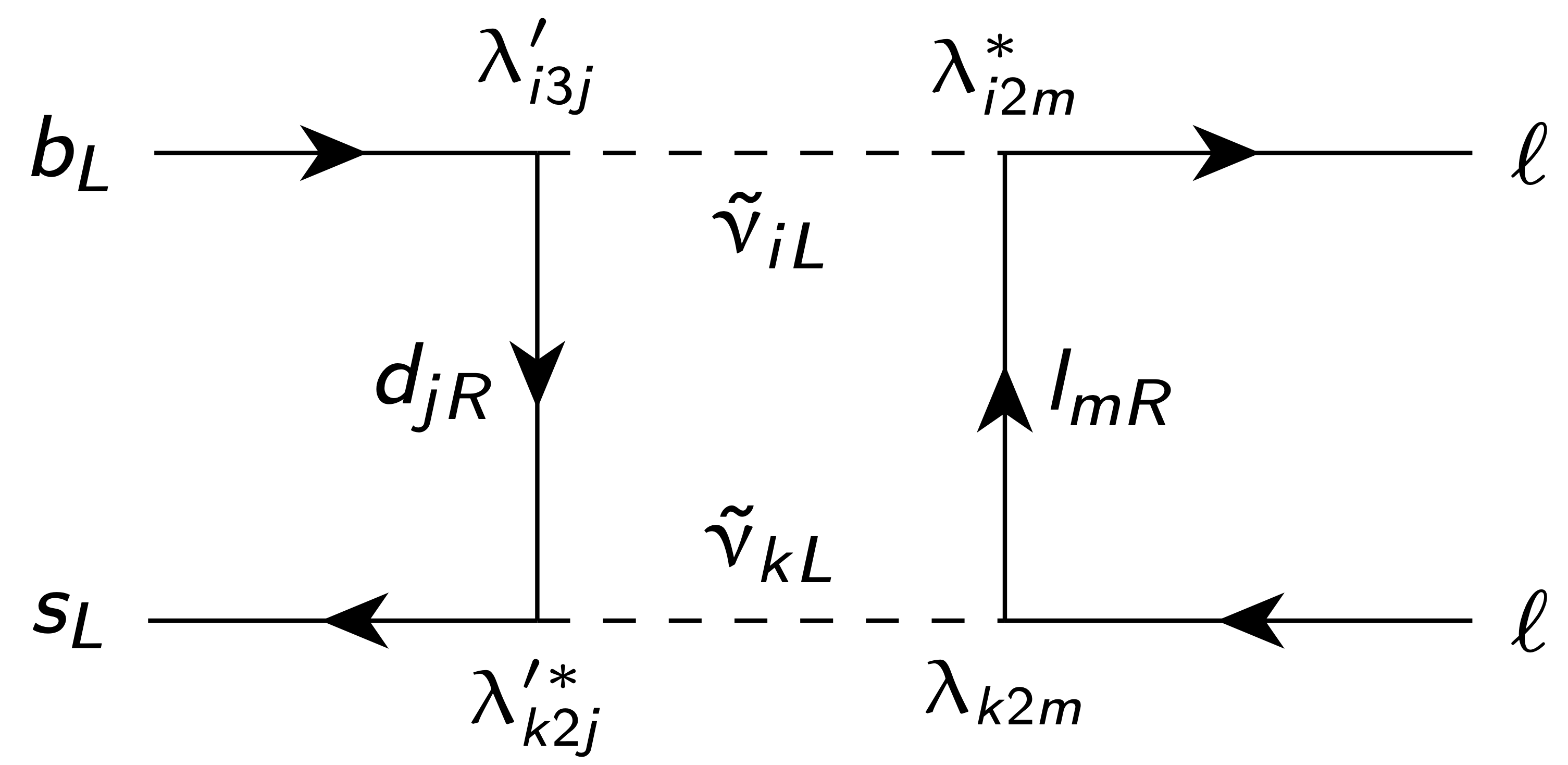}
\includegraphics[scale=0.2]{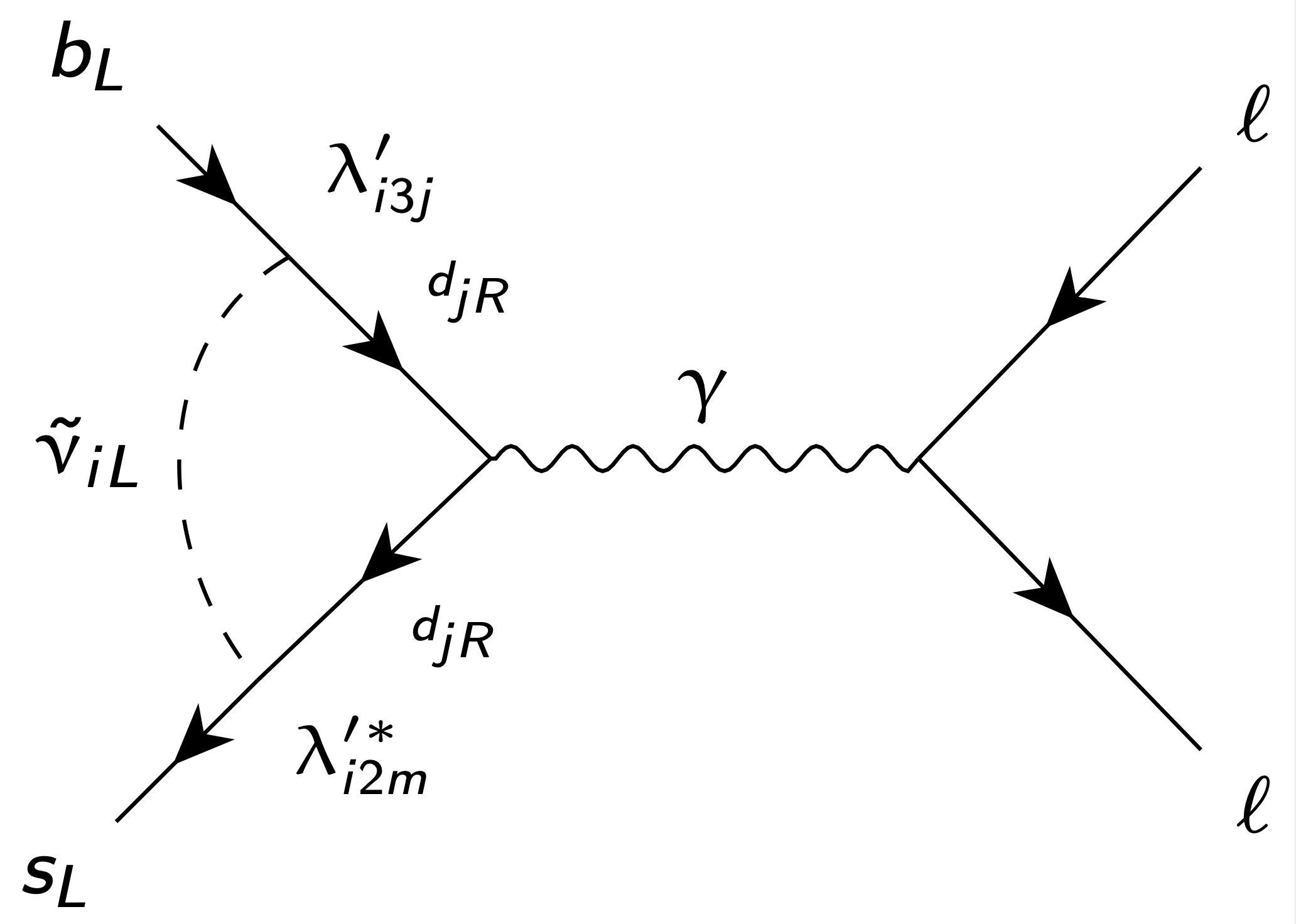}
\caption{The box and penguin Feynman diagrams contributing to $C_9^\ell$ and $C_{10}^\ell$.}
\label{fig:bsll_sneutrino}
\end{figure}

The mediation of $\tilde{\nu}$ can contribute to $b\to s\mu^+\mu^-$ processes at loop level via the photonic penguin diagram and a box-diagram, shown in Fig.~\ref{fig:bsll_sneutrino}. After integrating out $\tilde{\nu}$, the contributions to the $B$-meson decays can be described by shifts in the Wilson coefficients of effective operators in the effective Hamiltonian:
\begin{eqnarray}
    \mathcal H_\text{eff} \ &=& \ - \frac{4 G_F}{\sqrt{2}} V_{ts}^* V_{tb} \frac{\alpha}{4\pi} \left[C_S^\ell O_S^\ell + C_7 O_7 + \sum_{i = 9,10} (C_i)^\ell (O_i)^\ell\right]\nonumber
    \label{eq:Heff}
\end{eqnarray}
with
\begin{align}
\mathcal{O}_{S(S')}^{\ell}&=\ (\bar sP_{R(L)} b)(\bar \ell \ell) \,, \label{eq:QS}\\
\mathcal{O}_{7(7')}&=\frac{m_b}{e}\left(\bar{s} \sigma_{\mu \nu} P_{R(L)} b\right) F^{\mu \nu} \\
 O_{9{(9')}}^\ell \ & = \ (\bar s \gamma_\alpha P_{L(R)} b)(\bar \ell \gamma^\alpha \ell) \,, \label{eq:Q9}\\
 O_{10{(10')}}^\ell \ & = \ (\bar s \gamma_\alpha P_{L(R)} b)(\bar \ell \gamma^\alpha \gamma_5 \ell) \,. 
 \label{eq:Q10} 
\end{align}
In the SM, the value of the coefficients are~\cite{Altmannshofer:2008dz}: 
\begin{eqnarray}
(C_9^\ell)_{\rm SM} \ &=& 4.211; \ \ \ (C_{10}^\ell)_{\rm SM} \ = \  -4.103 \nn \\
 (C_7)_{\rm SM} &=& -0.304 ; \ \ \  (C_S^\ell)_{\rm SM} \simeq 0 
\end{eqnarray}
In order to explain the anomaly, using global fits, the New Physics contribution to $C_9$ and $C_{10}$ needs to be~\cite{Angelescu:2021lln} (see also Ref.~\cite{Alguero:2021anc,Geng:2021nhg,Carvunis:2021jga}):
\begin{equation}
\delta C_9^\mu = -\delta C_{10}^\mu = -0.41 \pm 0.09
\label{eqn:c9_c10_explanation}
\end{equation}

 We look at the different contributions to these Wilson coefficients and important constraints one-by-one. Our 
goal is to explain the requirement laid down in Eqn.~\ref{eqn:c9_c10_explanation} without violating any of the constraints. 

{\it Scalar Operators:}
 The scalar operator $O_{S}^\ell$ contributes to the $b\to s\ell^+\ell^-$ process at the tree level. It induces the
following change in the scalar Wilson coefficient:
\begin{equation}
 \delta C_{S}^{\mu}\,=\,\frac{\sqrt{2}\pi}{G_F\alpha V^\ast_{ts}V_{tb}}\frac{\sum_i\lambda^{\prime\ast}_{i23}\lambda_{i22}}{m_{\tilde{\nu}^2}}
 \label{eqn:scalar_op_wc}
\end{equation}
It is known that the scalar operators involving muons are unable to provide solutions to these anomalies (refer to Ref.~\cite{Ghosh:2017ber}). Moreover, the branching ratio of $B_s\to \mu^+\mu^-$ puts a strong constraint on the Wilson coefficient of the scalar operators. Thus, either the coupling constant in Eqn.~\ref{eqn:scalar_op_wc} is extremely small or the mass of the sneutrino is very large. We assume that the (sum of the) couplings is very small. 

{\it The contribution to $C_{9^\prime}$ and $C_{10^\prime}$:}
As discussed earlier, global fits of all the relevant data on rare $B$ decays as well as $\rkrks$ favor a 
BSM picture which is characterized by $C_9^\mu\, = -C_{10}^\mu$. This result also requires BSM in the 
primed coefficients to be subdominant. The present scenario, however, generates contributions to 
$C_{9^\prime}^{\ell}\, = -C_{10^\prime}^\ell$ due to mediation of $\tilde{u}_L$ at tree level and
 $\tilde{\nu}_L$ at loop level (Box diagram). These lead to an anti-correlated effect in $R_K$ and $R_{K^*}$~\cite{Ghosh:2017ber} and it is necessary to avoid this contribution. There are two ways to implement this: i) by postulating that only one generation of the right-handed quarks can couple to the other two fields, i.e. there is only one value of $k$ in $\lambda'_{ijk}$~\cite{Earl:2018snx,Das:2017kfo}, or, ii) by postulating that the squarks are extremely heavy and don't contribute to this process. We already have chosen squarks to be heavy because of which the tree-level contribution is very small. 
 To avoid the box diagram contribution, the first option is imperative. Therefore, we utilize the first option and choose $k=3$.
 
This leads to an important consequence. The tree-level contribution to $B_s - \bar{B}_s$ via sneutrino
exchange goes to zero, since the coupling involved in this process is of the form $\sum_i \lambda^{\prime \ast}_{i32} \lambda^\prime_{i23}$. This is zero since all $\lambda^\prime_{i32}$ are
zero consistent with the postulate above. 

 {\it The Box and Penguin contribution to $C_9$ and $C_{10}$:}
The contributions of the box~\cite{Altmannshofer:2020axr} and penguin~\cite{Hu:2019ahp} diagrams, shown in Fig.~\ref{fig:bsll_sneutrino}, to $C_9^\ell$ and $C_{10}^\ell$ Wilson coefficient are given by : 
\begin{eqnarray}
\delta C_9^{\ell, {\rm Box}} &=& \,\frac{\sqrt{2}\pi\,\lambda^\prime_{i33} \lambda^{\prime*}_{k23} \lambda^*_{i\ell m} \lambda_{k\ell m}}{8\,i\,G_F\alpha\,V_{tb}V^*_{ts}}\,D_2[m_{\tilde{\nu}}^2,m_{\tilde{\nu}}^2,m_d^2,m_\ell^2] \nn\\
\,&=&\,-\frac{\sqrt{2}}{128\,\pi\,G_F\alpha\,V_{tb}V^*_{ts}}\frac{\sum_{i,k,m}\lambda^\prime_{i33} \lambda^{\prime*}_{k23} \lambda^*_{i\ell m}\lambda_{k\ell m}}{m_{\tilde{\nu}}^2} \nn \\
&=& - \delta C_{10}^\ell  \\
\delta C_9^{\ell, {\rm penguin}}\,&=&\,-\frac{\sqrt{2}\sum_{i}\lambda^\prime_{i33}\lambda^{\prime \ast}_{i23}}{36G_F V_{tb}V^*_{ts} m_{\tilde{\nu}}^2}\Big[\frac{4}{3}+\log\Big(\frac{m_b^2}{m_{\tilde{\nu}}^2}\Big)\Big] \nn \\
&=& - \delta C_{10}^\ell  \label{eqn:peng_c9}
\label{eqn:c9c10}
\end{eqnarray}
where $\ell \,=\,e,\mu,\tau (\equiv 1,2,3)$ and $D_2[...]$ is the four-point Passarino-Veltman function for which the external momenta have been ignored. Note that the penguin contribution is lepton universal as opposed to the box diagram. Allowed the proper value of the couplings, these contributions should be able to explain the $\rkrks$ anomaly. The constraints on the value of the couplings come primarily from two sources -- the $b \to s \gamma$ process and the $B_s - \bar{B}_s$ oscillations.

{\it Photonic penguin contribution to $C_7$}:
The penguin diagram also leads to the effective operator $O_7$. The contribution to the $C_7$ coefficient is given by~\cite{Hu:2019ahp}: 
\begin{eqnarray}
\delta C_7^{\rm penguin}\,&=&\,\frac{\sqrt{2}\sum_{i} \lambda'_{i33}\lambda^{\prime \ast}_{i23} }{144 G_F V_{tb}V^*_{ts} m_{\tilde{\nu}}^2} \label{eqn:peng_c7}
\end{eqnarray}
 Given that the branching ratio of the process $b\to s \gamma$ is extremely well-measured by experiments, it provides useful limits on the contribution of the penguin diagram to the $C_7$ coefficient. This becomes all the more important since the RPV coupling in Eqn.~\ref{eqn:peng_c7} is the exact coupling which appears in Eqn.~\ref{eqn:peng_c9}.

From the world average of the measurement of $\mathcal{B}(B \to X_s \gamma)$, we have from HFAG \cite{Amhis:2019ckw}
\begin{equation}
\mathcal{B}\left( B \to X_s \gamma \right)|_{\rm exp} = (3.27 \pm 0.14) \times 10^{-4}
\end{equation}
(for $E_\gamma > E_0 = 1.6\, {\rm GeV}$)
while the SM prediction is \cite{Misiak:2015xwa}
\begin{equation}
\mathcal{B}\left( B \to X_s \gamma \right)|_{\rm SM} = (3.36 \pm 0.23) \times 10^{-4}
\end{equation}
This tight constrains leaves little room for NP contributions to $C_7$. According to Ref.~\cite{Paul:2016urs},
the real part of $C_7$ can be modified by 
\begin{eqnarray}
\delta C_7 &\in& [-0.018, 0.012]   \ \ {\rm at} \ \  1\sigma  \nn\\
&\in& [-0.032, 0.027]\ \  {\rm at} \ \  2\sigma
\label{eqn:c7_bound}
\end{eqnarray}
From this, we get the following constraint:
\beq
-1.81 \times 10^{-6}\,\leq\,\frac{\sum_{i} \lambda'_{i33}\lambda^{\prime \ast}_{i23} }{(m_{\tilde{\nu}}/{\rm GeV})^2}\leq 2.15 \times 10^{-6}
\label{eqn:C7bound}
\eeq

Based on various collider searches of RPV-SUSY models in which the sneutrino is the NLSP and the $\tilde{\chi}_1^0$ is the LSP, 
we assume $m_{\tilde{\nu}}\,=\,1~{\rm TeV}$ \cite{Aad:2021qct}. Note that the value of $m_{\tilde{\nu}}$ occurs only in the logarithm term of Eqn.~\ref{eqn:peng_c9} and thus the bound on $\delta C_9^{\rm penguin}$ is not sensitive to the exact value of $m_{\tilde{\nu}}$ (see conclusion for more details on the masses of sneutrinos). Using the $1\sigma$ bound from Eqn.~\ref{eqn:C7bound} in Eqn.~\ref{eqn:peng_c9}, the contribution of $\delta C_9^{\ell,\rm penguin}$ becomes
\beq
-1.68\,\leq\,\delta C_9^{\ell, {\rm penguin}}\,\leq\,1.41
\label{eqn:c9_bound_photonic_penguin}
\eeq

\begin{figure}
\centering
\includegraphics[scale=0.3]{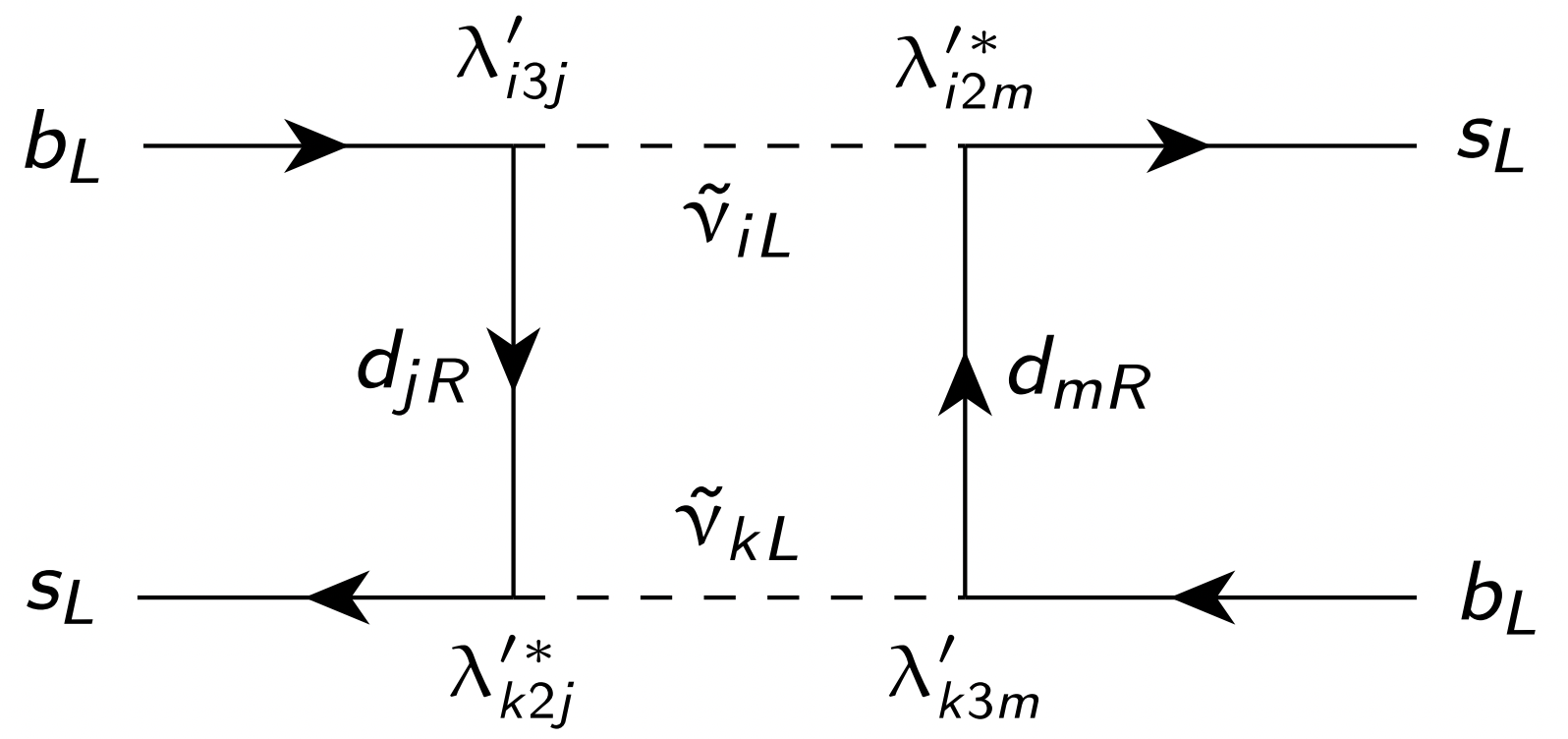}
\caption{The Feynman diagram for $B_s-\bar{B}_s$ mixing.}
\label{fig:bsbs_sneutrino}
\end{figure}
{\it Constraint from $B_s - \bar{B}_s$ mixing}:
The box diagram also contributes to $B_s - \overline{B}_s$ mixing as shown in Fig.~\ref{fig:bsbs_sneutrino}. Following the UTFit prescription \cite{Alpigiani:2017lpj}, we can parameterize the the full (SM + NP) oscillation amplitude to be: 
\begin{equation}
C_{B_s} e^{2 i \phi_{B_s}} = \frac{\langle B_s^0 \left| H_{\rm eff}^{\rm full}\right| \overline{B}_s\rangle}{\langle B_s^0 \left| H_{\rm eff}^{\rm SM}\right| \overline{B}_s\rangle}
\label{eqn:cbs}
\end{equation}

The RPV contribution can then be written as
\begin{equation}
C_{B_s} e^{2 i \phi_{B_s}} = 1 + \frac{\sum_{ik}\lambda^\prime_{i33}\lambda^{\prime \ast}_{k23} \lambda^{\prime \ast}_{i23}\lambda^\prime_{k33}}{32 M_W^2 G_F^2 S_0(x_t) \left(V_{tb} V_{ts}^\ast\right)^2\,m_{\tilde{\nu}}^2} 
\end{equation}
where $S_0(x)$ is the Inami-Lim function, which arises from the SM contribution, and $x_t~=~(m_t^2 / M_W^2)$, where $m_t$ and $M_W$ are the masses of the top and the $W$-boson respectively. We take $m_t =  172.76 {\rm \ GeV}$, we get $S_0(x_t) = 2.5264$. From this we get, 
\begin{eqnarray}
C_{B_s} &=& \left| 1 + \frac{\sum_{ik}\lambda^\prime_{i33}\lambda^{\prime \ast}_{k23} \lambda^{\prime \ast}_{i23}\lambda^\prime_{k33}}{32 M_W^2 G_F^2 S_0(x_t) \left(V_{tb} V_{ts}^\ast \right)^2\,m_{\tilde{\nu}}^2} \right| \nn \\
&=& 1 + 8.796 \times 10^6 \frac{\sum_{ik}\lambda^\prime_{i33}\lambda^{\prime \ast}_{k23} \lambda^{\prime \ast}_{i23}\lambda^\prime_{k33}}{(m_{\tilde{\nu}}/{\rm GeV})^2}
\label{eqn:BsBs}
\end{eqnarray}
The constraint on the value of $C_{B_s}$, defined in Eqn.~\ref{eqn:cbs} following the UTfit collaboration prescription \cite{Alpigiani:2017lpj}, is $C_{B_s} = 1.110 \pm 0.090$ and $\phi_{B_s} = (0.42 \pm 0.89) ^\circ$. Given that the SM values of the parameters are $C_{B_s} = 1, \phi_{B_s} = 0^\circ$, the NP contribution should be $C_{B_s}^{\rm NP}\,=\, 0.11\pm 0.09$. To be conservative, we work with only upper limit and thus, we get the following constraints on 
\beq
\left(\frac{\sum_{i}\lambda^\prime_{i33}\lambda^{\prime \ast}_{i23}}{(m_{\tilde{\nu}}/{\rm GeV})^2}\right)^2 \,\leq\, \frac{2.27\times 10^{-8}}{(m_{\tilde{\nu}}/{\rm GeV})^2}
\eeq
\beq
-\frac{1.5\times 10^{-4}}{(m_{\tilde{\nu}}/{\rm GeV})}\,\leq\,\frac{\sum_{i}\lambda^\prime_{i33}\lambda^{\prime \ast}_{i23}}{(m_{\tilde{\nu}}/{\rm GeV})^2} \,\leq\, \frac{1.5\times 10^{-4}}{(m_{\tilde{\nu}}/{\rm GeV})}
\label{eqn:Bs_bound}
\eeq
For $m_{\tilde{\nu}}\,=\, 1\,{\rm TeV}$, the constraint arising from $B_s-\bar{B}_s$ mixing measurement is much stronger than that from the measurement of $\mathcal{B}(b\to s\gamma)$\footnote{Strictly speaking, this situation will be realized if any one generation is involved, not necessarily the third.}. For $m_{\tilde{\nu}}\,=\,1~{\rm TeV}$, we use Eqn.~\ref{eqn:Bs_bound} in Eqn.~\ref{eqn:peng_c9} to get the following constraint on $\delta C_9^{\ell, {\rm penguin}}$ from $B_s-\bar{B}_s$ mixing:
\beq
-0.12\,\leq\,\delta C_9^{\ell, {\rm penguin}}\,\leq\,0.12 
\label{eqn:C9_bound_penguin_Bs}
\eeq
which far supersedes the weaker constraint derived in Eqn.~\ref{eqn:c9_bound_photonic_penguin}. We should  treat it as a phenomenological constraint which should always be satisfied by the corresponding combination of $\lambda'$ couplings. However, this contribution is lepton flavor universal and leads to non-zero $\delta C_9^{e,\tau}$ as well. Thus, to prevent a contribution to $C_{9}^{e}$,  consistent with our assumption, we postulate  the relation among the $\lambda^\prime$ couplings to follow the condition:
\begin{equation}
\sum_i \lambda^\prime_{i33} \lambda^{\prime \ast}_{i23} \simeq 0\,.
\label{eqn:rpv_gim}
\end{equation}
In this way, the contributions from penguin diagrams to $C_9$ (Eqn.~\ref{eqn:peng_c9}) and $C_7$ (Eqn.~\ref{eqn:peng_c7}) vanish (if the sneutrinos are degenerate in mass), leaving only the contribution from the box diagram. In view of Eqn.~\ref{eqn:rpv_gim}, the NP contribution to $b\to s \bar{\nu}\nu$ processes, via exchange of sneutrinos and $Z-$boson in the penguin diagram, also disappears. Potential constraints on the couplings can also arise from the box diagram involved in $D^0 - \bar{D}^0$ oscillations; however, the coupling combination involved in that process is $\left|\lambda^\prime_{i13} \lambda^\prime_{j23}\right|^2$, which is not the combination we consider. The couplings involved in the two processes are not correlated, thus, we do not consider this 
constraint in our analysis.

\begin{figure}
\includegraphics[scale=0.55]{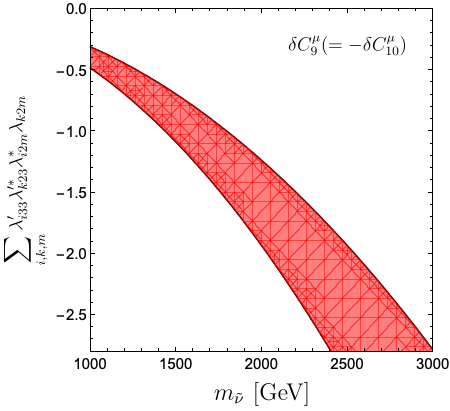}
\caption{Graph showing the couplings and the mass of the sneutrino required so that the contributions from the box diagram can explain the $\rkrks$ anomaly. It is assumed that the entire contribution to $\delta C_9^\mu$ is due to the Box diagram. }
\label{fig:rk_explanation}
\end{figure}

\begin{table*}[t]
\centering
\begin{tabular}{|lcc|}
\hline
{\bf Process} & {\bf RPV coupling} & {\bf Constraint on coupling } \\
 & {\bf involved} & {\bf so that contribution vanishes}\\
\hline
$b\to c \tau \nu$ & $\lambda^{\prime \ast}_{i23} \lambda_{ij3}$ & \multirow{2} {*}{$\lambda_{ij3} = 0$} \\
$b\to u \tau \nu$ & $\lambda^{\prime \ast}_{i13} \lambda_{ij3}$ & \\
\hline
$\tau \to \mu \gamma$ & $\lambda_{ij3}\lambda_{ij2}^\ast$ & Zero because of $\lambda_{ij3} = 0$ \\
& $+\,\lambda_{i3k}\lambda_{i2k}^\ast$ & $\lambda_{ij2} = 0$ and $\lambda_{132} = 0$ \\ 
\hline
$\mu \to e \gamma$ & $ \lambda_{ij2} \lambda^\ast_{ij1}$  &   Zero because of $\lambda_{ij2} = 0$ \\ 
& $+\, \lambda_{i2j} \lambda^\ast_{i1j} $ &$\lambda_{ij2} = 0$ and $\lambda_{131} = 0$  \\
\hline
$Z \to e_{R} \mu_{R}$ & $\lambda_{ij1} \lambda^\ast_{ij2}$ & Zero because of above choices \\ 
$Z \to e_{R} \tau_{R}$ &  $\lambda_{ij1} \lambda^\ast_{ij3}$ &  \\
$Z \to \mu_{R} \tau_{R}$ & $\lambda_{ij2} \lambda^\ast_{ij3}$ &  \\
\hline
$Z \to e_{L} \mu_{L}$ & $\lambda_{i1j} \lambda^\ast_{i2j}$ & Zero because of above choices \\ 
$Z \to e_{L} \tau_{L}$ &  $\lambda_{i1j} \lambda^\ast_{i3j}$ & With $\lambda_{ij2} = 0$;  $\lambda_{211}\lambda_{231}\lesssim 71.0 $(for $m_{\tilde{\nu}}=1\,{\rm TeV}$) \\
$Z \to \mu_{L} \tau_{L}$ & $\lambda_{i2j} \lambda^\ast_{i3j}$ & Zero because of above choices\\
\hline
\end{tabular}
\caption{Summary of the low energy constraints relevant to the couplings involved in the $R_{K^{(*)}}$ 
process. While the explicit limits on the couplings have been calculated in the Appendix, the table lists
out the condition on the couplings such that the contribution to the respective processes is zero. Note
that this always leaves $\lambda^\prime_{i33} \lambda^{\prime \ast}_{k23} \lambda_{i22} \lambda^\ast_{k22}$ non-zero, which contributes to $b\to s \mu^+ \mu^-$ and thus to $\rkrks$. } 
\label{tab:low_energy_couplings} 
\end{table*}

\begin{figure}
\includegraphics[scale=0.48]{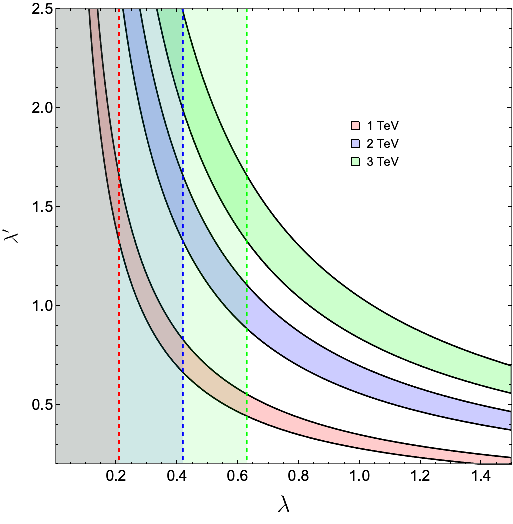}
\caption{Contour region on $\lambda-\lambda^\prime$ coupling explaining the $\rkrks$ anomaly corresponding to assumptions made in Eqn.~\ref{eqn:BP2} for different values of $m_{\tilde{\nu}}$. The
bands represent the regions required for the explanation of $\rkrks$ at $1\sigma$ C.L., including
the global fit constraints. The shaded region left of vertical lines is allowed by $\mu\to e\gamma$ process measurment, which provides the strongest constraints.}
\label{fig:BP2}
\end{figure}
{\it Explaining the $\rkrks$ anomaly}: Using the assumption in Eqn.~\ref{eqn:rpv_gim}, the values of the couplings and the mass of the sneutrinos that explain $\rkrks$ within $1\sigma$ are plotted in Fig.~\ref{fig:rk_explanation}. From this, we conclude that the box diagram alone can provide a phenomenologically viable explanation for both the $\rk$ and $\rks$ anomalies \footnote{It has to be emphasized here that this solution {\it also} takes into account other constraints like those arising from $B_s - \bar{B}_s$
and $B_s \to \mu^+ \mu^-$ process. This is because the best fit value of the $\delta C_9$ coefficient comes from global fits, which takes into account these constraints.}.

There are certain conditions and caveats, however, which need to be stated. Firstly, our solution requires at least two generations of slepton doublets, unlike in Ref.~\cite{Altmannshofer:2020axr} where only the third generation slepton doublet was considered. Secondly, we require $\delta C_9^{e, box}=0$, which can be achieved by setting the corresponding coupling to zero. Doing this doesn't affect the contribution of the box diagram to the $\delta C_9^{\mu, box}$ coefficient. Thirdly, from Fig.~\ref{fig:rk_explanation}, it is clear that in large parts of the parameter space which explain the anomalies, the value of the couplings is large which means that many of the corresponding $\lambda$ and $\lambda^\prime$ couplings might themselves be $\mathcal{O}(1)$. This can potentially be a problem as these couplings also contribute to various decays such as $\ell_i\to\ell_j\gamma$, $B\to\ell_i\nu_j,\,B\to\ell_i\ell_j$ etc. However, we have checked that the constraints on the couplings arising from these processes can be easily accommodated by keeping certain couplings to be small, while still explaining the $\rkrks$ anomaly. These constraints are discussed in detail in the Appendix. 
A summary of the low energy constraints are given in Table~\ref{tab:low_energy_couplings}.

{In order to illustrate the fact that  the low energy constraints do not affect our explanation
of $\rkrks$, we can consider two benchmark scenarios. 

{\it Case 1: Constrained benchmark} :
In this scenario, the RPV contribution to all low energy processes is zero and this can be achieved if the conditions outlined in the third column of Table~\ref{tab:low_energy_couplings} are all assumed. Even in this highly constrained scenario, the coupling product in Eqn.~\ref{eqn:c9c10} becomes 
\begin{eqnarray}
{\textstyle\sum_{i,k,m}}\lambda^\prime_{i33} \lambda^{\prime*}_{k23} \lambda^*_{i\ell m}\lambda_{k\ell m}&\sim&\lambda^{\prime\ast}_{123}\lambda_{121}(\lambda^\prime_{133}\lambda^\ast_{121}+\lambda^\prime_{333}\lambda^\ast_{321}) \nonumber\\
&+&\lambda^{\prime\ast}_{323}\lambda_{321}(\lambda^\prime_{133}\lambda^\ast_{121}+\lambda^\prime_{333}\lambda^\ast_{321})\nonumber\\
\end{eqnarray}
According to Ref.~\cite{Rakshit:1998kd}, $\lambda_{321}$ and $\lambda_{121}$ can have magnitude less than unity. So, we can safely assume the product, $\lambda_{321}.\lambda_{121}$, to be small in comparison to $|\lambda_{321}|^2$ or $|\lambda_{121}|^2$, just for our convenience and we get
\beq\textstyle\sum_{i,k,m}\lambda^\prime_{i33} \lambda^{\prime*}_{k23} \lambda^*_{i\ell m}\lambda_{k\ell m}
\sim\lambda^{\prime\ast}_{123}\lambda^\prime_{133}|\lambda_{121}|^2+\lambda^{\prime\ast}_{323}\lambda^\prime_{333}|\lambda_{321}|^2.\eeq
Note that by keeping the product, $\lambda_{321}\lambda_{121}$ non-negligible would increase the required parameter space. Thus, our approach is conservative.

For $\tilde{m}_\nu\,\sim\,1\,{\rm TeV}$, we can choose 
\begin{eqnarray}
\lambda^\prime_{133}\sim\,0.02,&\quad& \lambda^\prime_{123}\sim\,3.5\nonumber\\
\lambda^\prime_{333}\sim\,3.5,&\quad& \lambda^\prime_{323}\sim\,3.5
\end{eqnarray}
with squark masses fixed around 100 TeV. $\lambda^\prime_{133}$ coupling contributes to Majorana mass for the neutrinos, which provides a very strong constraint. The coupling scales as the square-root of the down type squark mass \cite{Rakshit:1998kd}. Since it's value at $m_{\tilde{d}} = 100\ {\rm GeV}$ is $\lambda^\prime_{133} \sim 7 \times 10^{-4}$ \cite{Bhattacharyya:1997vv}, at $m_{\tilde{d}} = 100\ {\rm TeV}$, we have $\lambda^\prime_{133} \sim 10^{-2}$. Note that the LQD couplings ($\lambda^\prime$) like $\lambda^\prime_{123}, \lambda^\prime_{323}$ and $\lambda^\prime_{333}$ at large squark masses are unconstrained, so we take the maximum value allowed by perturbative unitarity (i.e. $\sqrt{4 \pi} \approx 3.5$). With this, we require $0.16\leq\lambda_{321}\leq0.21$ to explain $\rkrks$ anomaly which is allowed by studies of perturbative unitarity \cite{Bhattacharyya:1997vv, Rakshit:1998kd}.} { Also, we assume the sign of $\delta C_9$ is taken care by the phase difference of difference $\lambda^\prime_{ijk}$ couplings.} {Thus, we can explain the $\rkrks$ anomalies in this benchmark, even if all other couplings are set to zero so that there is no RPV contribution to the low energy processes.

{\it Case 2: Less constrained benchmark}: 
We can also illustrate another scenario in which we don't force all the low energy constraints to be zero.
In this scenario, we assume the following: 
\begin{eqnarray}
\lambda_{ij1}\,&=&\,10^{-2}\lambda\quad \lambda_{ij2}\,=\,\lambda \nonumber \\ 
{\rm with}&\quad& \lambda_{ij3}\,=\,0,\quad \lambda_{31j}\,=\,0 \nonumber\\ 
{\rm and}&\quad&\lambda^\prime_{233}\,=\,\lambda^\prime_{333}\,=\,\lambda^\prime_{i23}\,=\,\lambda^\prime \nonumber\\
{\rm with}&\quad& \lambda^\prime_{133}\,=\,10^{-2}
\label{eqn:BP2}
\end{eqnarray}}
{Note that the above assumptions correspond to real part of couplings. For the imaginary part, we assume that the phase difference in various $\lambda^\prime_{ijk}$ couplings explains sign of $\delta C_9$.} {We choose $\lambda_{ij3}\,=\,0$ to evade strong constraints from $b\to\,c\tau\nu$ whereas $\lambda_{ij1}$ should be small in comparison to $\lambda_{ij2}$ to avoid $\mu\to e\gamma$ limits. We keep $\lambda_{31j}$ so small that we can avoid bounds from $\mu_L\to\,e_L\gamma$, $\tau_L\to\mu_L\gamma$, $Z\to\,e_L\mu_L$ and $Z\to\,\mu_L\tau_L$. To this end, we get contour regions on $\lambda-\lambda^\prime$ plane addressing $\rkrks$ anomaly for different sneutrino masses in Fig.~\ref{fig:BP2}. Note that, for this case, the relevant low energy constraints are due to the $\mu_R \to e_R \gamma$ $Z\to e_R \ \mu_R$ and $Z\to e_L \tau_L$ processes. Out of these, the strongest constraint 
is due to $\mu_R \to e_R \gamma$, which is what is plotted.  

Both of the results above imply that it is possible to evade the low energy constraints while still continuing to explain $\rkrks$ in our scenario.
}

\section{Conclusion}

Chinks in the seemingly impregnable armor of the Standard Model (SM) are rare, but one of the most promising ones appear from the measurement of the ratio of the branching ratios of semi-leptonic decays of the $B$-mesons, like $R_{D^{(*)}}$ or $R_{K^{(*)}}$. A recent measurement of the ratio $\rk$ by the {\sc LHCb} collaboration announced at the 2021 Moriond Conference
strengthens the existing discrepancy in $R_K$ with the SM from $2.5\sigma$ earlier to $3.1 \sigma$ now. Several theoretical attempts have been made to explain this within some
extension of the SM. We attempt the same using a simplified RPV-SUSY framework, using only lepton number-violating couplings, in which all particles except left-handed slepton doublet are too heavy to significantly 
contribute to the $b \to s \ell \ell$ ($\ell = e, \mu$) process. We show that even with this minimal extension, it is possible to explain the anomaly with the 
box diagram contribution alone, while also respecting all relevant constraints. We stress that our proposed solution is by no means
the only solution to the problem, but that it is a phenomenologically plausible one. Specific values of couplings are considered which are compatible with these assumptions, without delving into an explanation of their particular value.

The various limits derived on the RPV-SUSY couplings and thus on the relevant Wilson coefficients of the effective operators crucially depend on the mass of the sneutrino. Collider searches in the RPV-SUSY scenario
where the $\tilde{\nu}$ is the NLSP and the $\tilde{\chi}^0_1$ is the LSP put the limits at $\sim 1$~TeV\cite{Aad:2021qct}. Additionally, the values of the $\lambda$ and $\lambda^\prime$ couplings are dependent on and scales appropriately with the mass of the squarks. Since the squarks in our minimal model are very heavy, there is no upper limit on the values of the RPV couplings, apart from those arising possibly from perturbative unitarity. More stringent lower bounds on sneutrino masses are obtained from studying the direct production of the $\tilde{\nu}$ in $p p$ collisions, which then decay exclusively to lepton flavour violating channels such as $e \mu$, $\mu \tau$ and $e \tau$. Collider searches for this signal in Ref.~\cite{Aaltonen:2010fv,Khachatryan_2016,Aaboud:2016hmk,Aaboud:2018jff} assume sneutrinos to be degenerate and rules out their existence for mass less than 3.4 TeV, 2.9 TeV, 2.6 TeV for tau sneutrino decaying to $e\mu, e\tau,$ or $\mu\tau$ respectively, with $\lambda^{\prime}_{i11,i22}\,=\,0.11$ and $\lambda_{i12,i13,i23..}\,=\,0.07$. Note,  however, that these analyses assume $Br(\tilde{\nu}\to\ell\ell^\prime)  =  1.0$. If sneutrino is kinematically allowed to decay overwhelmingly into a neutralino ($\tilde{\chi}^0_1$) and SM neutrino, the aforementioned constraints gets relaxed and limits on $m_{\tilde{\nu}}$ comes to be about $1~{\rm TeV}$~\cite{ATLAS:2018rns,ATLAS:2014pjz}. While we have used the more relaxed mass limit in our calculations, it is perfectly possible to use the larger mass limit and still arrive at a plausible explanation. 

\begin{acknowledgments}
DG and DS acknowledge support through the Ramanujan Fellowship and MATRICS Grant of the Department of Science and Technology, Government of India. The project has received funding (in part) from the European Union's Horizon 2020 research and innovation programme under grant agreement No. 101002846, ERG CoG "CosmoChart". The research of DB was supported in part by the Israel Science Foundation (grant no.  780/17), the United States - Israel Binational Science Foundation (grant no.2018257) and by the Kreitman Foundation Post-Doctoral Fellowship. 
\end{acknowledgments}

\appendix
\section{Tree-level contribution to $b\to c \tau \nu$}
The RPV coupling that goes into the $\rkrks$ ratios is the combination:
\begin{equation}
\rkrks \propto \lambda^\prime_{i33} \lambda^{\prime \ast}_{k23} \lambda_{i2m} \lambda^\ast_{k2m}
\end{equation} 
where $i, k \neq 2$ owing to the antisymmetry in the first  two indices of the $\lambda$ couplings. 
\begin{equation}
\boxed{\mathbf{Assume:}\ \ \quad \lambda_{ij3} = 0}
\label{eqn:rpv_assume}
\end{equation}

This is to make the tree-level contributions to $b\to c/u \ \tau \nu$ or $B_{u/c} \to \tau \nu$ vanish.  The amplitude of this process is proportional
to the RPV coupling
\begin{equation}
\mathcal{A}_{tree}^{b \to c} \propto \frac{\lambda^\prime_{ik3} \lambda_{ij3}}{m_{\tilde{l_i}}^2}
\end{equation}
which vanishes with the assumption in Eqn.~\ref{eqn:rpv_assume}. 

\section{Charged Lepton Flavour Changing Processes}
RPV couplings contribute to processes of the type $\ell_i \to \ell_j \gamma$. We shall explore these and
try to calculate a bound on the contributions. 
\begin{figure}
\includegraphics[scale=0.15]{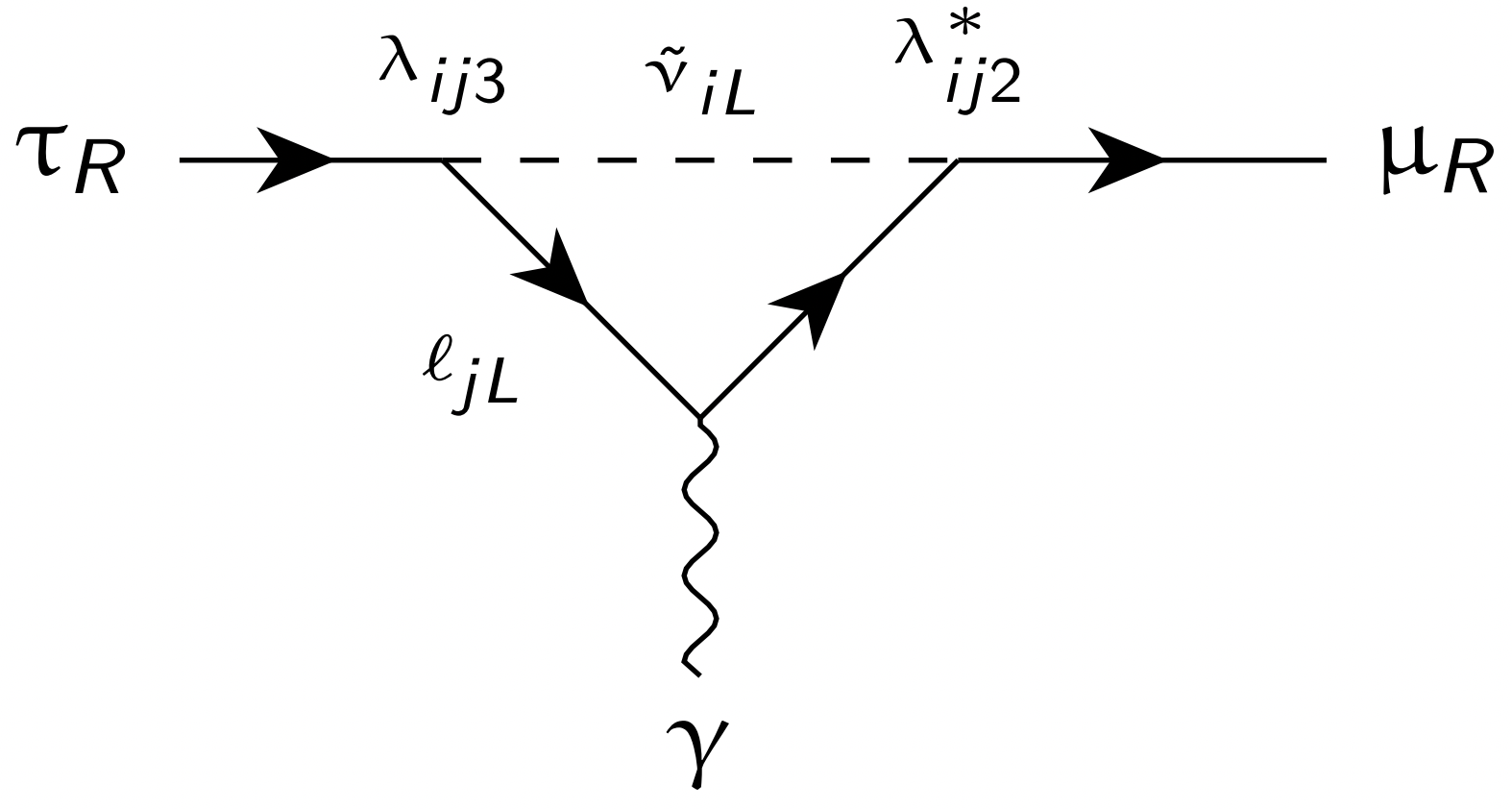}
\includegraphics[scale=0.15]{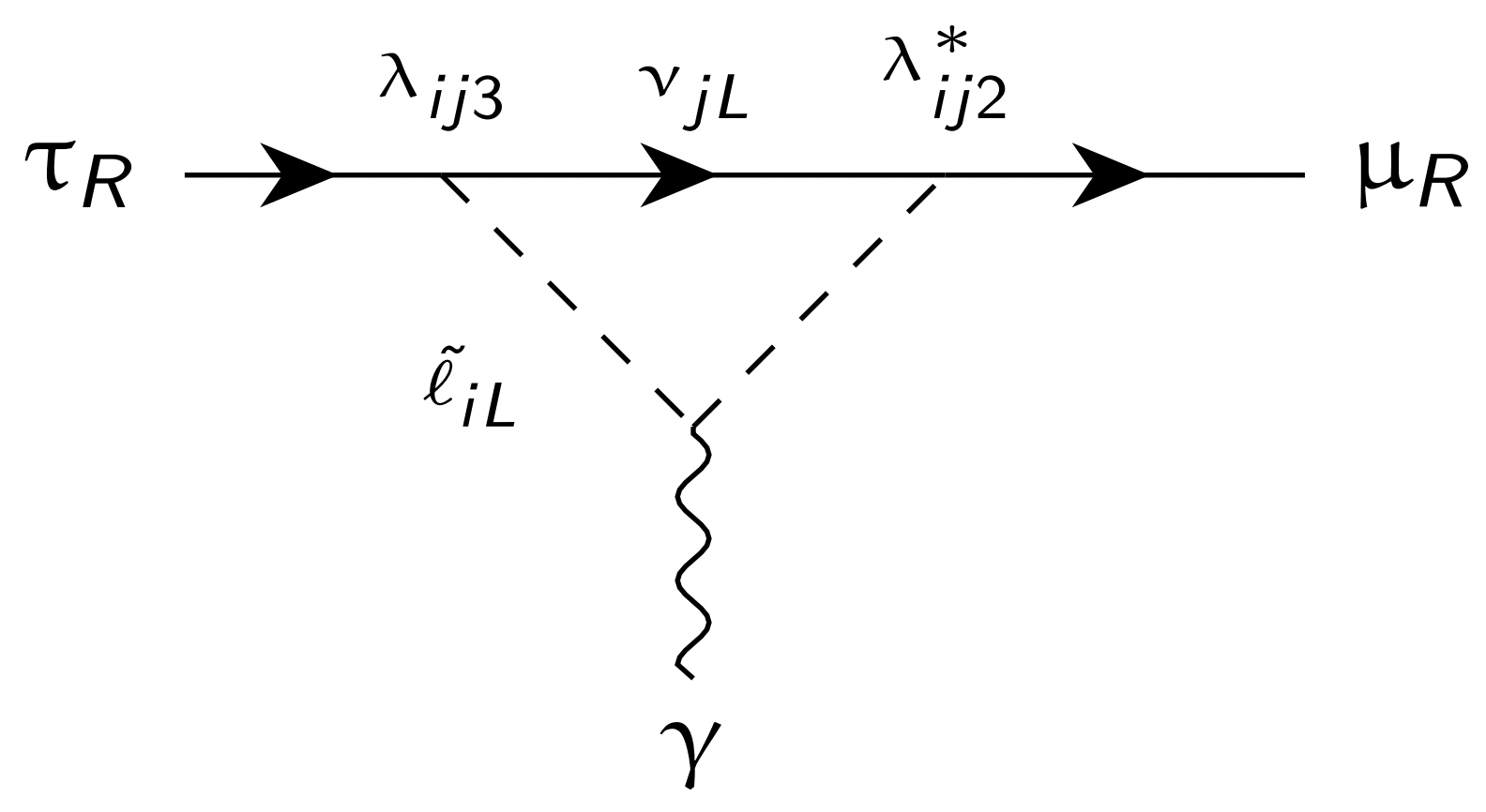}
\includegraphics[scale=0.15]{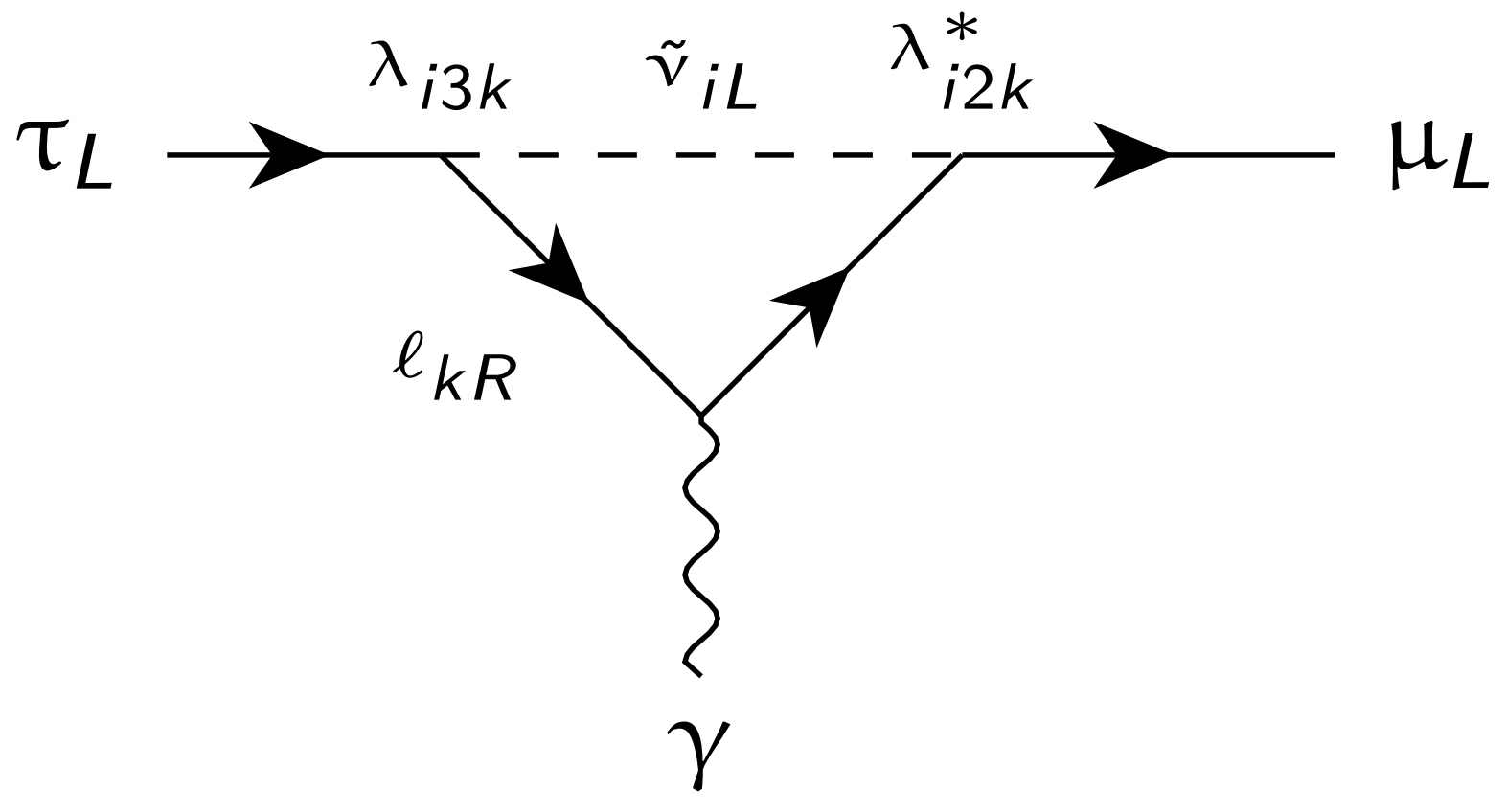}
\includegraphics[scale=0.15]{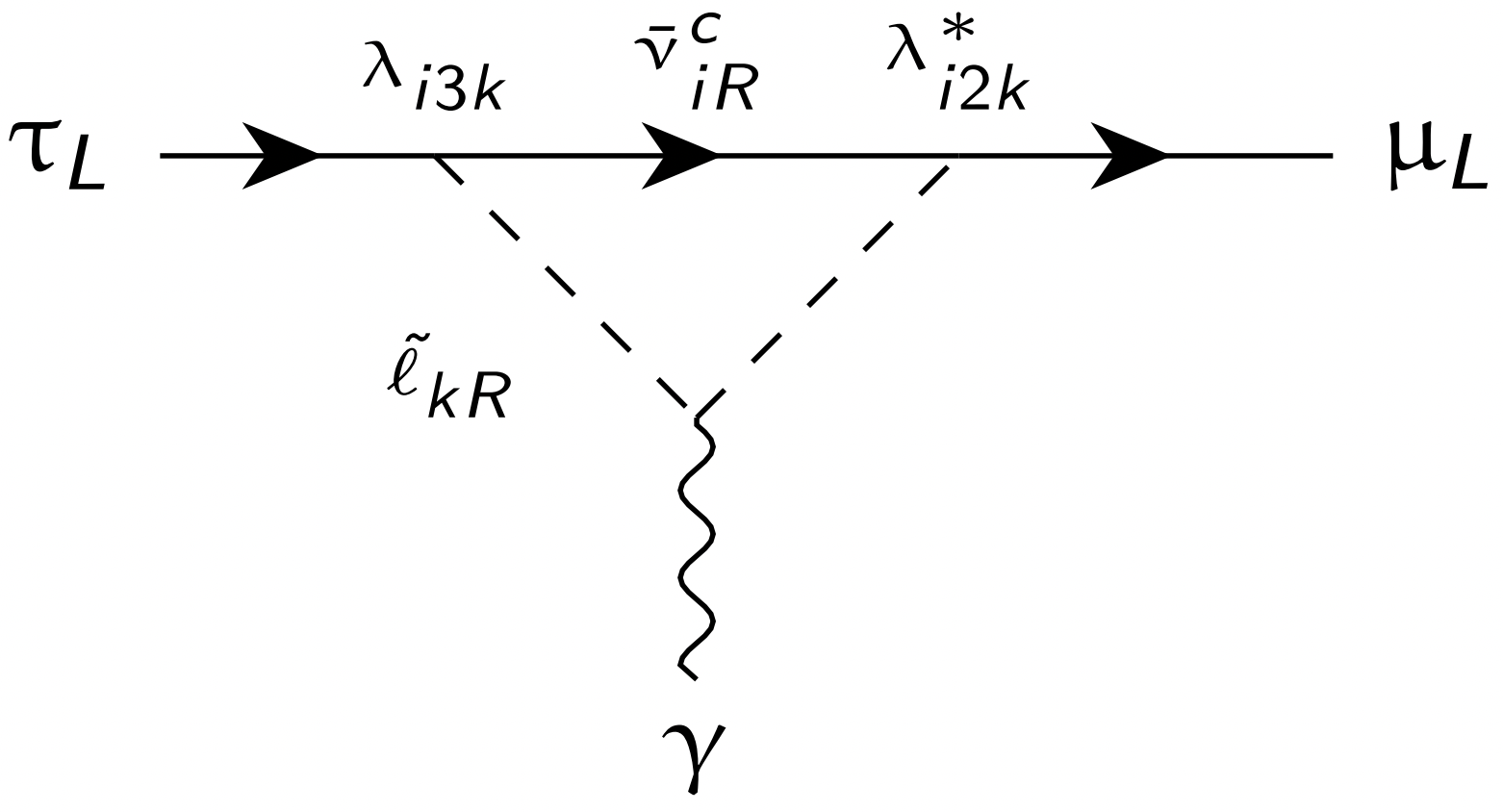}
\caption{Feynman diagrams showing the process $\tau \to \mu \gamma$ for both handedness
of the external leptons. The diagrams for $\mu \to e \gamma$ and $\tau \to e \gamma$ are similar
with appropriate changes of the couplings.}
\label{fig:liljgamma}
\end{figure}
\subsection{$\tau \to \mu \gamma$ process}
The process $\tau \to \mu \gamma$ can involve both right handed and left handed charged leptons. 
We have to treat them separately, since the RPV couplings involved in them are different. 

\noindent
{\bf Right handed current:}\\
The Feynman diagrams for this process are given on the top row of Fig.~\ref{fig:liljgamma}

The amplitude for these diagrams are proportional to
\begin{eqnarray}
\mathcal{A}_{\tau_R \to \mu_R \gamma}^{(1)} &\propto& \frac{\lambda_{ij3}\lambda_{ij2}^\ast}{m_{\tilde{l_{Li}}}^2} \\
\mathcal{A}_{\tau_R \to \mu_R \gamma}^{(2)} &\propto& \frac{\lambda_{ij3}\lambda_{ij2}^\ast}{m_{\tilde{\nu_{Li}}}^2} 
\end{eqnarray}
Evidently both of these contributions vanish due to the assumption in Eqn.~\ref{eqn:rpv_assume}. Thus,
for the process $\tau\to\mu \gamma$ there is no contribution from right-handed charged leptons. \\

\noindent
{\bf Left handed current:} \\
The diagrams for this process are given in the bottom row of Fig.~\ref{fig:liljgamma}.
For left-handed leptons, there is a non-zero contribution. The amplitude is proportional to
\begin{eqnarray}
\mathcal{A}_{\tau_L \to \mu_L \gamma}^{(1)} &\propto & \frac{\lambda_{i3k}\lambda_{i2k}^\ast}{m_{\tilde{l}_{Rk}}^2}\\
\mathcal{A}_{\tau_L \to \mu_L \gamma}^{(2)} &\propto & \frac{\lambda_{i3k}\lambda_{i2k}^\ast}{m_{\tilde{\nu}_{Li}}^2}
\end{eqnarray}

The RPV coupling, written explicitly, is
\begin{eqnarray}
\lambda_{i3k}\lambda_{i2k}^\ast &=& \lambda_{i31}\lambda_{i21}^\ast  + \lambda_{i32}\lambda_{i22}^\ast \nonumber \\
&=& \lambda_{131}\lambda_{121}^\ast  + \lambda_{132}\lambda_{122}^\ast
\end{eqnarray}
Note that whereas the value of $k$ in the first  line cannot be 3 because of Eqn.~\ref{eqn:rpv_assume}, 
the value of $i$ in the second line can only be 1 because of the anti-symmetry in the first two indices 
of the $\lambda$ coupling. 

The decay width of  this process then is
\begin{eqnarray}
\Gamma \left(\tau \to \mu \gamma\right)|_{\rm RPV} \simeq \frac{\alpha_{em} m_\tau^5}{256 \pi^4} \kappa^2 = 5.186 \times 10^{-6} \kappa^2 ~{\rm GeV}^5 \nn \\
\end{eqnarray}
where 
\begin{eqnarray}
\kappa & = & \frac{1}{m_{\tilde{l}}^2} \left( \lambda_{131}\lambda_{121}^\ast  + \lambda_{132}\lambda_{122}^\ast \right) \equiv \frac{\mathcal{C}(\lambda)}{m_{\tilde{l}}^2}
\end{eqnarray}
For a slepton or sneutrino mass of about 1 TeV, this gives, 
\begin{equation}
\Gamma \left(\tau \to \mu \gamma\right)|_{\rm RPV} =  5.19 \times 10^{-18} |\mathcal{C}(\lambda)|^2 ~ {\rm GeV}
\end{equation}
The total decay width is $\Gamma(\tau \to \mu \gamma) = 4.0 \times 10^{-13}~{\rm GeV}$. 
Thus, the branching ratio is given by
\begin{equation}
\mathcal{B} \left(\tau \to \mu \gamma\right)|_{\rm RPV} =  1.30 \times 10^{-5} |\mathcal{C}(\lambda)|^2 
\end{equation}
The experimental limit on the branching ratio is $\mathcal{B} \left(\tau \to \mu \gamma\right) < 4.4 \times 10^{-8}$~\cite{ParticleDataGroup:2020ssz}. This means that we can avoid this bound if $|\mathcal{C}(\lambda)| < 0. 06$. This
is quite achievable with the current limits on the RPV couplings. We can however completely 
eliminate this contribution by postulating that the only non-zero LLE RPV couplings are of the 
form $\lambda_{i22}$, since this is what is required for our explanation of $R_{K^{(\ast)}}$.

\subsection{$\mu \to e \gamma$ process}
As with the previous process, this can also involve both right and left handed charged leptons. The diagrams are similar and the amplitudes are proportional to 
\begin{eqnarray}
\mathcal{A}_{\mu_R \to e_R \gamma}^{(1)} &\propto & \frac{\lambda_{ij2} \lambda^\ast_{ij1}}{m_{\tilde{\nu}_{Li}}^2} \\
\mathcal{A}_{\mu_R \to e_R \gamma}^{(2)} &\propto & \frac{\lambda_{ij2} \lambda^\ast_{ij1}}{m_{\tilde{\l}_{Lj}}^2} \\
\mathcal{A}_{\mu_L \to e_L \gamma}^{(1)} &\propto & \frac{\lambda_{i2j} \lambda^\ast_{i1j}}{m_{\tilde{\nu}_{Rj}}^2} \\
\mathcal{A}_{\mu_L \to e_L \gamma}^{(2)} &\propto & \frac{\lambda_{i2j} \lambda^\ast_{i1j}}{m_{\tilde{\l}_{Rj}}^2} 
\end{eqnarray}

The contribution to the decay width of the process due from RPV-SUSY is given by
\begin{equation}
\left. \Gamma (\mu \to e \gamma)\right|_{\rm RPV} \simeq \frac{\alpha_{\rm em} m_\mu^5}{256 \pi^4} \kappa^2 = 4.0 \times 10^{-12} \kappa^2 \ {\rm GeV}^{5}
\end{equation}
where, given $m_{\tilde{\nu}} \simeq m_{\tilde{l}}$, 
\begin{eqnarray}
\kappa &=& \frac{1}{m_{\tilde{l}}^2}  \left[\lambda_{i2j} \ \lambda_{i1j}^\ast - \lambda_{ij2}\ \lambda_{ij1}^\ast \right] \nn \\
&=& \frac{1}{m_{\tilde{l}}^2} \left[\lambda_{321} \ \lambda_{311}^\ast + \lambda_{322} \ \lambda_{312}^\ast - \lambda_{122}\ \lambda_{121}^\ast - \lambda_{132}\ \lambda_{131}^\ast \right. \nn \\
&& \left. \quad - \lambda_{212}\ \lambda_{211}^\ast -\lambda_{232}\ \lambda_{231}^\ast - \lambda_{312}\ \lambda_{311}^\ast - \lambda_{322}\ \lambda_{321}^\ast \right]\nn \\
&=& \frac{1}{m_{\tilde{l}}^2} \left[\lambda_{321} \ \lambda_{311}^\ast + \lambda_{322} \ \lambda_{312}^\ast -  2 \lambda_{122}\ \lambda_{121}^\ast - 2 \lambda_{232}\ \lambda_{231}^\ast \right. \nn \\
&& \left. \quad - 2 \lambda_{312}\ \lambda_{311}^\ast \right] \label{eqn:muegamma_coupling}\\
&\equiv &\frac{\mathcal{C}(\lambda)}{m_{\tilde{l}}^2}
\end{eqnarray} 
Considering the mass of the sleptons $m_{\tilde{l}} \simeq 1 \ {\rm TeV}$, we have
\begin{equation}
\left. \Gamma (\mu \to e \gamma)\right|_{\rm RPV} \simeq 4.0 \times 10^{-24} \left| \mathcal{C}(\lambda) \right|^2 \ {\rm GeV}
\end{equation}. 
Given that the total decay width is $\Gamma_\mu = 3.0 \times 10^{-19}$~GeV, this gives us a 
branching ratio of: 
\begin{equation}
\mathcal{B}(\mu \to e \gamma) \simeq 1.33 \times 10^{-5} \left| \mathcal{C}(\lambda) \right|^2
\end{equation}

The limit on the branching ratio from the MEG collaboration is $\mathcal{B}(\mu \to e \gamma) < 4.2 \times 10^{-13}$ \cite{MEG:2016leq}. This puts strong bounds on the possible values of the
coupling. Naively, one can say that each of the lambda couplings is $\mathcal{O}(10^{-2})$. Looking
at the couplings involved (Eqn.~\ref{eqn:muegamma_coupling}), we find that  only the $\lambda_{122}$ 
coupling needs to be large for our explanation of $R_{K^{(\ast)}}$. We can easily postulate that the only
couplings which are non-zero are of the form $\lambda_{i22}$. 
This will be enough to explain the anomalies in our scenario. This postulate will also mean that the 
value of $\kappa$ in Eqn.~\ref{eqn:muegamma_coupling} is zero, which forbids this decay via RPV 
couplings. 

\section{$Z \to \ell_i \ell_j$ }
The one-loop diagrams for the decay of the $Z$-boson to dileptons (including those of different flavours) in the RPV-SUSY framework are given by: 
\begin{figure}
\includegraphics[scale=0.15]{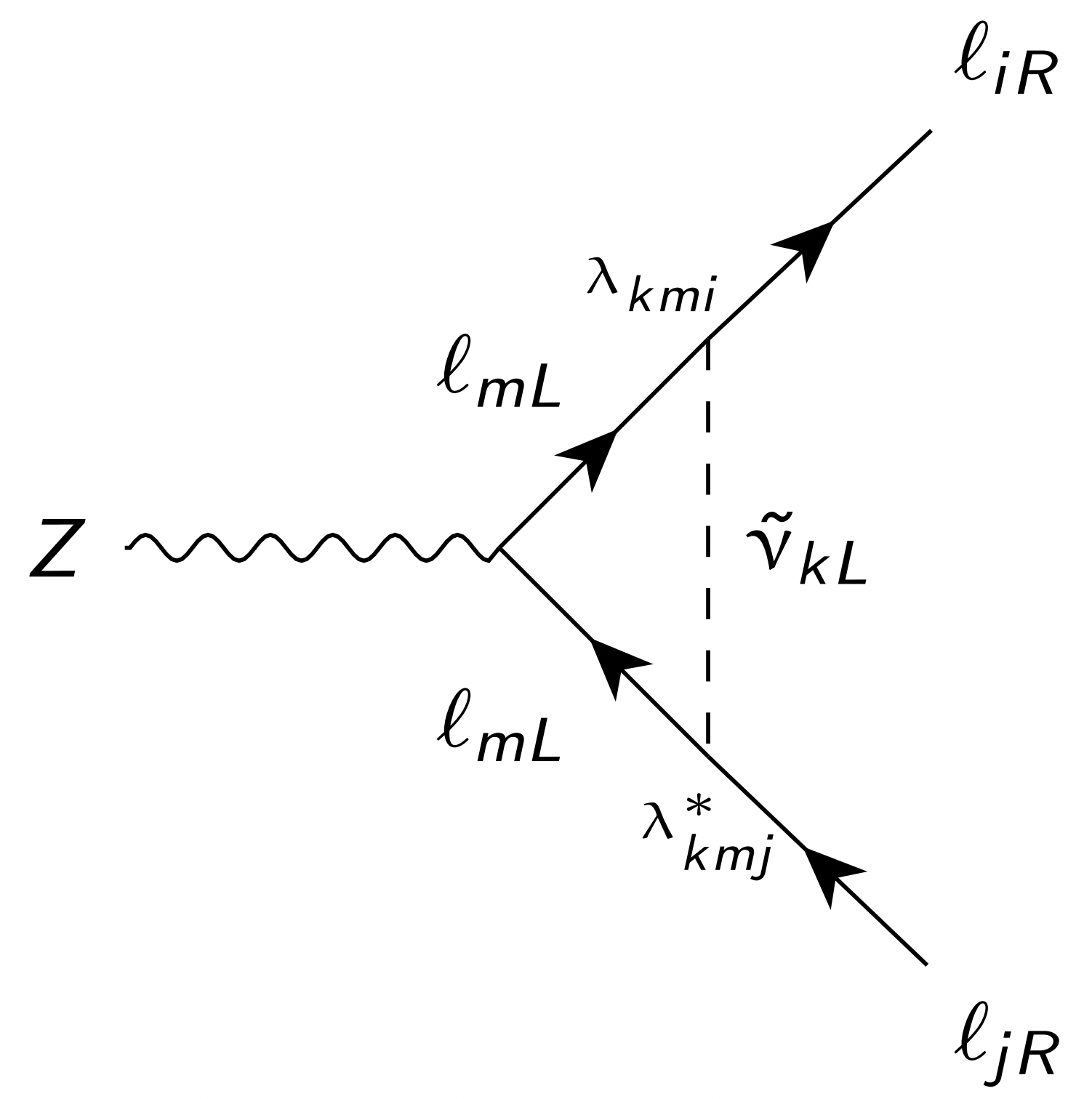}
\includegraphics[scale=0.15]{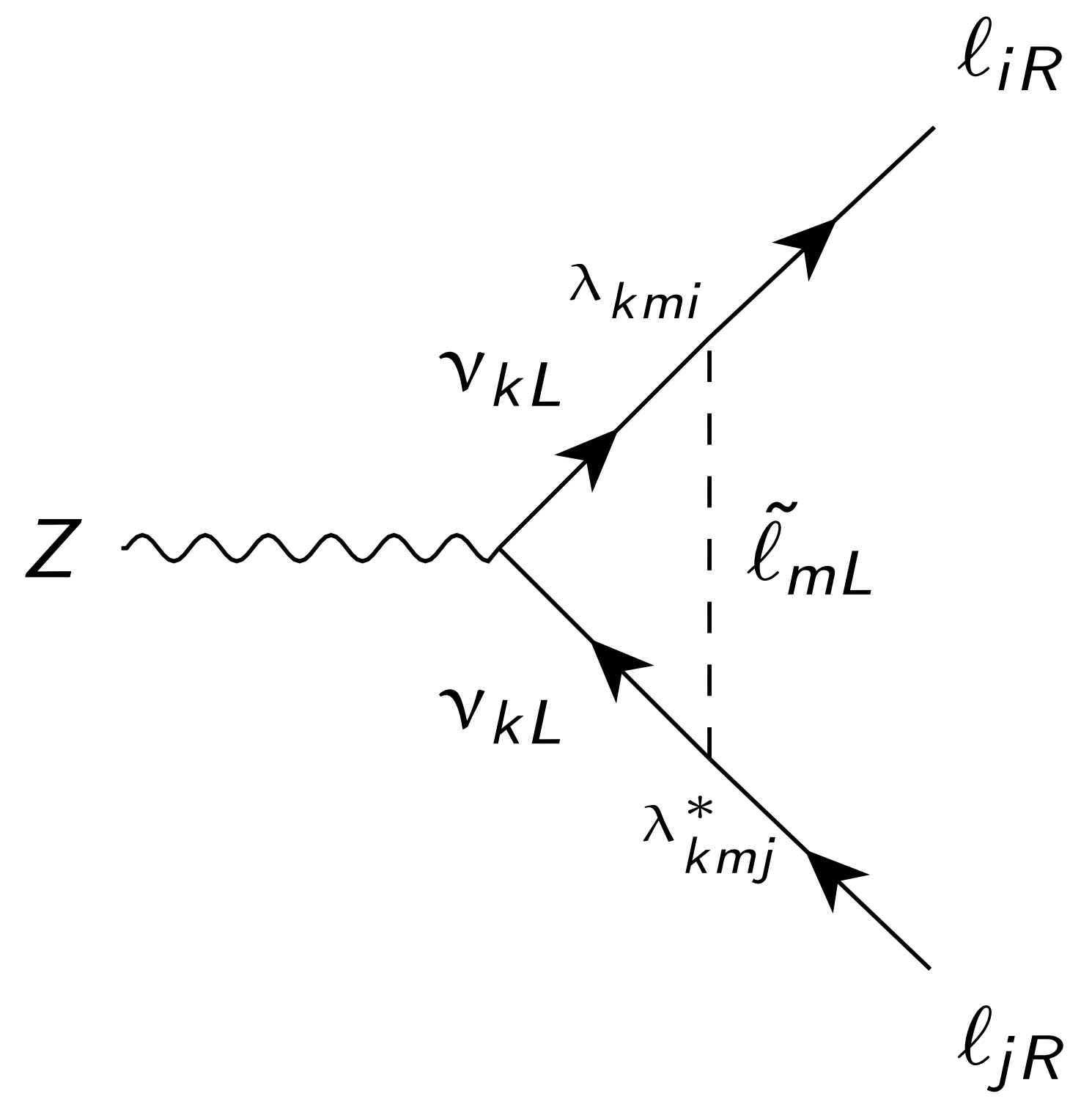}
\includegraphics[scale=0.15]{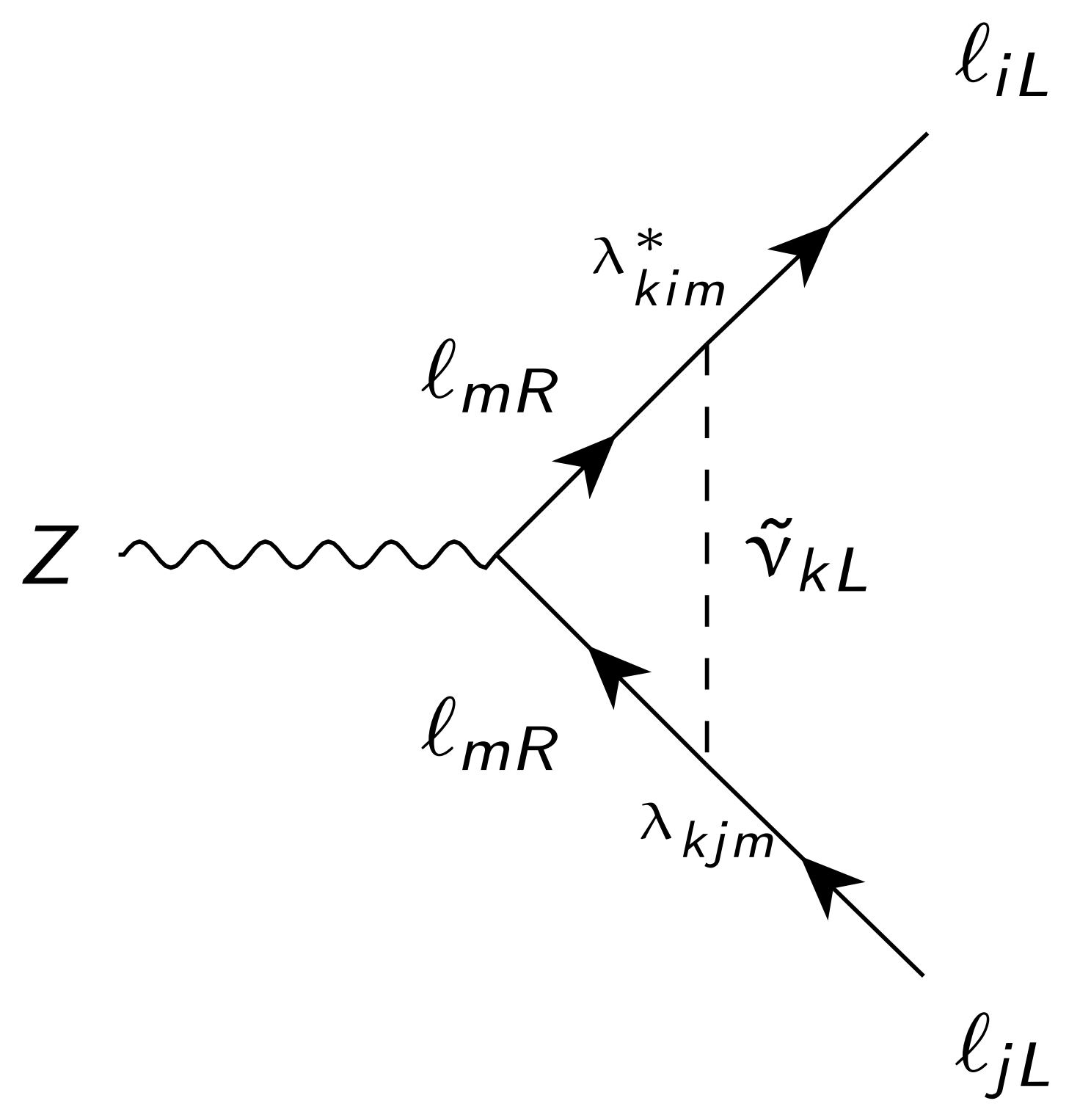}
\caption{Feynman diagrams for the process $Z \to \ell_i \ell_j$ via RPV-SUSY scenario, involving only sleptons and sneutrinos.}
\label{fig:Zll_Feynman}
\end{figure}

For right-handed final state leptons, the matrix element can be written as 
\begin{eqnarray}
\mathcal{M} &=& \bar{u}(p_2) \lambda^\ast_{knj} \int \frac{d^D q}{(2\pi)^D} \frac{1}{q^2 - m_{\tilde{\nu}_k}^2} \frac{\slashed{q} - \slashed{p}_2-m_{l_m}\mathbb{1}}{(q-p_2)^2 - m_{l_m}^2} \nn \\ 
&& \times \ \ g_L \gamma^\mu P_L \frac{\slashed{q} + \slashed{p}_1-m_{l_m}\mathbb{1}}{(q+p_1)^2 - m_{l_m}^2} \lambda_{kni} v(p_1) \epsilon_\mu (p_1 + p_2) \nonumber \\
&=& g_L \lambda^\ast_{knj} \lambda_{kni} \epsilon_\mu (p_1 + p_2) \bar{u}(p_2)  \int \frac{d^D q}{(2\pi)^D} \frac{1}{q^2 - m_{\tilde{\nu}_k}^2} \nn \\ 
&& \times\ \  \frac{{\rm Tr_D}[(\slashed{q} - \slashed{p}_2-m_{l_m}\mathbb{1}) \gamma^\mu P_L (\slashed{q} + \slashed{p}_1-m_{l_m}\mathbb{1})]}{((q-p_2)^2 - m_{l_m}^2)((q+p_1)^2 - m_{l_m}^2)}    v(p_1) \nonumber
\end{eqnarray}
where ${\rm Tr_D}$ is the Dirac trace. Using Package-X \cite{Patel:2015tea} to evaluate the trace, we have
\begin{eqnarray}
\mathcal{M} &=& g_L \lambda^\ast_{knj} \lambda_{kni} \epsilon_\mu (p_1 + p_2) \bar{u}(p_2)  \nn \\
&& \int \frac{d^D q}{(2\pi)^D} \frac{1}{q^2 - m_{\tilde{\nu}_k}^2} \frac{-2 m_{m_l} \left(p_{1\mu} - p_{2\mu} -2q_\mu \right)}{((q-p_2)^2 - m_{l_m}^2)((q+p_1)^2 - m_{l_m}^2)}    v(p_1) \nonumber
\end{eqnarray}
Evaluating  the loop integral only gives us: 
\begin{eqnarray}
L &=& -2 m_{l_m} \left[ \left(p_1^\mu - p_2^\mu\right) C_0 \left(m_{\tilde{\nu}}, m_{l_m}, m_{l_m};  m_Z, m_{l_i}, m_{l_j}\right)  \right. \nn \\ 
&&\left.  \quad \quad +2 p_1^\mu C_1 \left(m_{\tilde{\nu}}, m_{l_m}, m_{l_m};  m_Z, m_{l_i}, m_{l_j}\right) \right. \nonumber \\
&& \left. \quad \quad - 2 p_2^\mu C_2\left(m_{\tilde{\nu}}, m_{l_m}, m_{l_m};  m_Z, m_{l_i}, m_{l_j}\right)\right] 
\label{eqn:Loop_func_Zll}
\end{eqnarray}
where 
\begin{eqnarray}
& C_{0}& \equiv  C_{0}\left(m_{1}, m_{2}, m_{3} ; M_{1}, M_{2}, M_{3}\right) =\\
&& \int  \frac{d^{4} k}{\pi^{2}} \frac{1}{\left(k^{2}+m_{1}^{2}\right)\left\{\left(k+p_{2}\right)^{2}+m_{2}^{2}\right\}\left\{\left(k+p_{2}+p_{3}\right)^{2}+m_{3}^{2}\right\}}  \nn\\
&  p_i^\mu C_{i}& \equiv  p_i^\mu C_{i}\left(m_{1}, m_{2}, m_{3} ; M_{1}, M_{2}, M_{3}\right)= \\ 
&& \int  \frac{d^{4} k}{\pi^{2}} \frac{q^\mu}{\left(k^{2}+m_{1}^{2}\right)\left\{\left(k+p_{2}\right)^{2}+m_{2}^{2}\right\}\left\{\left(k+p_{2}+p_{3}\right)^{2}+m_{3}^{2}\right\}}  \nn
\end{eqnarray}
 and ${\rm where}\ p_i^2 = -M_i^2; \ p_1 = p_2 + p_3)$

Evaluating the loop functions numerically for $M_Z~=~91.2\ {\rm GeV}, m_{l} = 1.776 \ {\rm GeV}$ and
$m_{\tilde{\nu}} \sim 1\ {\rm TeV}$, we have 
\begin{equation}
L = (1.79815 \times 10^{-6} + 1.53181 \times 10^{-8} i) (p_1 - p_2)^\mu \equiv \kappa (p_1-p_2)^\mu \nn
\end{equation}
Thus we have
\begin{eqnarray}
\mathcal{M} &=& g_L \kappa \lambda^\ast_{kmj} \lambda_{kmi} \epsilon_\mu (p_1 + p_2)  (p_1-p_2)^\mu \bar{u}(p_2) v(p_1) \nn
\end{eqnarray}
which gives
\begin{eqnarray}
|\mathcal{M}|^2 &=& \left| g_L \kappa \lambda^\ast_{kmj} \lambda_{kmi} \right|^2 \epsilon_\mu (p_1 + p_2) \epsilon^\ast_\nu (p_1 + p_2) \nn \\
&& (p_1-p_2)^\mu (p_1-p_2)^\nu {\rm Tr}[(\slashed{p}_2 + m_l)(\slashed{p}_1 - m_l)] \nn\\
&=& \left| g_L \kappa \lambda^\ast_{kmj} \lambda_{kmi} \right|^2 \epsilon_\mu (p_1 + p_2) \epsilon^\ast_\nu \nn \\
&& (p_1 + p_2) (p_1-p_2)^\mu (p_1-p_2)^\nu [4( p_1.p_2 - m_{l_m}^2)] \nn\\
&=& 2 \left(m_Z^2 - 4 m_{l_m}^2\right) \left| g_L \kappa \lambda^\ast_{kmj} \lambda_{kmi} \right|^2 \epsilon_\mu (p_1 + p_2)  \nn \\
&& \epsilon^\ast_\nu (p_1 + p_2) (p_1-p_2)^\mu (p_1-p_2)^\nu
\end{eqnarray}
Using the completeness relations for the massive Z-boson, we have
\begin{eqnarray}
|\mathcal{M}|^2 &=& 2 \left(m_Z^2 - 4 m_{l_m}^2\right) \left| g_L \kappa \lambda^\ast_{kmj} \lambda_{kmi} \right|^2 \nn \\
&& \left(-g_{\mu\nu} + \frac{(p_1 + p_2)_\mu (p_1 + p_2)_\nu}{M_Z^2}\right) (p_1-p_2)^\mu (p_1-p_2)^\nu \nn \\
&=& - 2 \left(m_Z^2 - 4 m_{l_m}^2\right)  \left| g_L \kappa \lambda^\ast_{kmj} \lambda_{kmi} \right|^2 (p_1-p_2). (p_1-p_2) \nn \\
&=& 2   \left(m_Z^2 - 4 m_{l_m}^2\right)^2 \left| g_L \kappa \lambda^\ast_{kmj} \lambda_{kmi} \right|^2 
\end{eqnarray}
Using $g_L = I_3 - Q \sin^2 \theta_{\rm W} = -\frac{1}{2} + \sin^2 \theta_{\rm W}$, we have
\begin{eqnarray}
|\mathcal{M}|^2 &=& 3.433 \times 10^{-5} \left|\lambda^\ast_{kmj} \lambda_{kmi} \right|^2
\end{eqnarray}

Thus, the decay width is
\begin{eqnarray}
\Gamma(Z \to \ell_i \ell_j)|_{\rm RPV} &=& \frac{|\mathcal{M}|^2}{16\pi\,m_Z}\sqrt{1-\frac{4m_{l_m}^2}{m_Z^2}} \\
&\simeq & 7.48 \times 10^{-9} \times \left|\lambda^\ast_{kmj} \lambda_{kmi} \right|^2 ~{\rm GeV}\nn 
\end{eqnarray}
The total decay width of the $Z$ boson is $\Gamma_Z \simeq 2.495$~GeV, so this gives us a branching
ratio of 
\begin{equation}
\mathcal{B}(Z \to \ell_i \ell_j)|_{\rm RPV} \simeq 3.0 \times 10^{-9} \times \left|\lambda^\ast_{kmj} \lambda_{kmi} \right|^2
\end{equation}
Comparing it to the experimental limits on the $Z$ boson decay to differently flavoured leptons at
95\% C.L., we have \cite{ParticleDataGroup:2020ssz}
\begin{eqnarray}
\mathcal{B}(Z \to e \mu )|_{\rm RPV} & < & 7.5 \times 10^{-7} \implies \left|\lambda^\ast_{km2} \lambda_{km1} \right| \lesssim 16.0 \nn \\
\mathcal{B}(Z \to e \tau )|_{\rm RPV} & < & 9.8 \times 10^{-6} \implies \left|\lambda^\ast_{km3} \lambda_{km1} \right| \lesssim 57.2\nn\\
\mathcal{B}(Z \to \mu \tau )|_{\rm RPV} & < & 1.2 \times 10^{-5} \implies \left|\lambda^\ast_{km3} \lambda_{km2} \right| \lesssim 63.2 \nn 
\end{eqnarray}

Owing to the assumption stated in Eqn.~\ref{eqn:rpv_assume}, the latter two branching ratios are zero
automatically. Using the antisymmetry of the first two indices, the coupling on the first line can be  expanded to
\begin{eqnarray}
\left| \lambda_{121} \lambda^\ast_{122} + \lambda_{131} \lambda^\ast_{132} +  \lambda_{231} \lambda^\ast_{232} \right| < 8.0  
\label{eqn:Zemu_limit}
\end{eqnarray}
Using the bounds outlined in Table II of Ref.~\cite{Kao:2009fg}, all the couplings mentioned here
are about $0.5 - 0.8$, assuming a right handed slepton mass of $\sim 1$~TeV. This clearly evades
the bounds in Eqn.~\ref{eqn:Zemu_limit}. 

The other diagram in Fig.~\ref{fig:Zll_Feynman} for $Z \to \ell_{iR} \ell_{jR}$ with neutrinos in the loop
vanishes, since the value of the loop function depends on the mass of the fermion in the loop, as can 
be seen from Eqn.~\ref{eqn:Loop_func_Zll}. The diagram for $Z \to \ell_{iL} \ell_{jL}$ gives an 
equivalent contribution to the diagram calculated above, given as:
\begin{eqnarray}
\mathcal{B}(Z \to e \mu )|_{\rm RPV} & < & 7.5 \times 10^{-7} \implies \left|\lambda^\ast_{km2} \lambda_{km1} \right| \lesssim 19.6 \nn \\
\mathcal{B}(Z \to e \tau )|_{\rm RPV} & < & 9.8 \times 10^{-6} \implies \left|\lambda^\ast_{km3} \lambda_{km1} \right| \lesssim 71\nn\\
\mathcal{B}(Z \to \mu \tau )|_{\rm RPV} & < & 1.2 \times 10^{-5} \implies \left|\lambda^\ast_{km3} \lambda_{km2} \right| \lesssim 78.6 \nn 
\end{eqnarray}

{Note that both of the above limits are obtained assuming Z boson decays to either left-handed leptons or right-handed leptons. If we assume both kind of leptons at once, the limits are weaker.}

\bibliography{reference}

\end{document}